\pdfoutput=1
\documentclass{article}
\usepackage{amsmath,amsthm,amssymb}
\usepackage[dvipsnames]{xcolor}
\usepackage{physics}
\usepackage{tensor}
\usepackage{graphicx}
\usepackage{float}
\usepackage{xcolor, colortbl}
\usepackage{hyperref}
\usepackage{authblk}
\usepackage{mathtools}
\definecolor{light-gray}{gray}{0.9} 
\usepackage{setspace}
\newcommand{\sltr}{\mathrm{SL}(2,\mathbb{R})}

\newenvironment{eqaed}
   {
    \begin{equation}
      \begin{aligned}
    }
    {
        \end{aligned}
    \end{equation}
    }
\title{Neural Networks Assisted Metropolis-Hastings for Bayesian Estimation of Critical Exponent on Elliptic Black Hole Solution in 4D Using Quantum Perturbation Theory}
 \date{}
\author[1]{Armin Hatefi\footnote{ahatefi@mun.ca}}
\author[2]{Ehsan Hatefi\footnote{ehsan.hatefi@uah.es}}
\author[3]{Roberto J. Lopez-Sastre\footnote{robertoj.lopez@uah.es}}
\affil[1]{Department of Mathematics and Statistics, Memorial University of Newfoundland, St. John’s, NL, Canada.\vspace{.01em}}
\affil[2,3]{University of Alcalá, Department of Signal Theory and Communications, Research group GRAM, Alcalá de Henares, Spain.\vspace{1em}}
\begin{document}
 \maketitle

\vspace{-1cm}
\begin{abstract}
It is well-known that the critical gravitational collapse produces continuous self-similar solutions characterized by the Choptuik critical exponent, $\gamma$. We examine the solutions in the domains of the linear perturbation equations, considering the numerical measurement errors. Specifically, we study quantum perturbation theory for the four-dimensional Einstein-axion-dilaton system of the elliptic class of $\text{SL}(2,\mathbb{R})$ transformations. We develop a novel artificial neural network-assisted Metropolis-Hastings algorithm based on quantum perturbation theory to find the distribution of the critical exponent in a Bayesian framework. 
Unlike existing methods, this new probabilistic approach identifies the available deterministic solution and explores the range of physically distinguishable critical exponents that may arise due to numerical measurement errors.
\end{abstract}

\section{Introduction}
Black holes are known to be the final state of gravitational collapse. They have a well-known characteristic property. They are entirely defined by mass, angular momentum, and charge. Choptuik \cite{Chop} shows that there seems to be another parameter or the fourth quantity that establishes the collapse itself. Following up on Christodoulou's work on the spherically symmetric collapse of scalar fields \cite{Christodolou}, Choptuik looked into the idea that a critical behaviour can show discrete spacetime self-similarity. The amplitude of the scalar field fluctuation shows that $p$ must be greater than the critical value $p_\text{crit}$ to form a black hole. Moreover, for values of $p$ above this threshold, the mass of the black hole $M_\text{bh}$ results a scaling law as
\begin{equation}
r_S(p) \propto M_\text{bh}(p) \propto (p-p_\text{crit})^\gamma\,.
\end{equation}

The critical exponent in 4d for a single real scalar field is given by $\gamma\simeq 0.37$~\cite{Chop,Hamade:1995ce,Gundlach:2002sx}, while for general dimension  ($d \geq 4$)~\cite{KHA,AlvarezGaume:2006dw} these relations are modified as
\begin{equation}
 r_S(p) \propto (p-p_\text{crit})^\gamma \,, \quad M_\text{bh}(p) \sim (p-p_\text{crit})^{(D-3)\gamma}.
\end{equation}

Various numerical simulations with other fields are carried out \cite{Birukou:2002kk,Husain:2002nk,Sorkin:2005vz,Bland:2005vu,HirschmannEardley,Rocha:2018lmv}. As an instance the collapse of a perfect fluid is done in~\cite{AlvarezGaume:2008qs,evanscoleman,KHA,MA}.
In \cite{evanscoleman} they explored $\gamma \simeq 0.36$ and therefore it was realised in \cite{Strominger:1993tt} that $\gamma$  might be universal for all matter field coupled to gravity in four dimensions. It was first in ~\cite{KHA,MA,Hirschmann:1995qx} that the authors found the critical exponent may be found by working out the perturbations of the solutions. To achieve the perturbations, one needs to perturb any field $h$ (be it the metric or the matter content) as 
\begin{equation}
    h = h_0 + \varepsilon \, h_{-\kappa},
\end{equation}
where the perturbation $h_{-\kappa}$ does have the scaling $-\kappa \in \mathbb{C}$ that labels the different modes. Within the possible values of $\kappa$, we express the most relevant mode $\kappa^*$ as the highest value of $\Re(\kappa)$. 

It was argued in ~\cite{KHA,MA,Hirschmann:1995qx} that the $\kappa^*$ is related to the critical exponent by $\gamma = \frac{1}{\Re \kappa^*}$. The axial symmetry was analyzed in \cite{AE}, and the critical solution for the shock waves was investigated by \cite{AlvarezGaume:2008fx}. 

For the first time, the axion-dilaton critical collapse solution coupled to gravity in four dimensions was determined by \cite{Hirschmann_1997} where they found the value $\gamma \simeq 0.2641$, hence arguing the doubts relating to the universality of $\gamma$ in four dimensions.
The first motivation for studying critical solutions in the axion-dilaton system is the AdS/CFT correspondence \cite{Maldacena:1997re}, which relates the Choptuik exponent to the imaginary part of quasinormal modes, and the dual conformal field theory \cite{Birmingham:2001hc}. Additional motivations include the holographic description of black hole formation \cite{AlvarezGaume:2006dw} and the broader physics of black holes and their applications to holography and string theory \cite{Hatefi:2012bp,Hatefi:2012wj}.

In type IIB string theory, there is significant interest in exploring gravitational collapse in spaces that can asymptotically approach $AdS_5 \times S^5$ where the matter content is described by the axion-dilaton system and the self-dual 5-form field. The paper \cite{ours} recently investigated entire families of distinguishable continuous self-similar solutions of the Einstein-axion-dilaton system in four and five dimensions, including all three conjugacy classes of $\sltr$. This study builds upon works of \cite{AlvarezGaume:2011rk, hatefialvarez1307}. By applying advanced analytic and numerical methods described in \cite{Antonelli:2019dqv}, the perturbed critical solution of a four-dimensional elliptic critical collapse was computed and determined the value of $\gamma$ to be around 0.2641, as previously reported by \cite{Hirschmann_1997}. The findings provide strong confidence in our ability to determine other critical exponents across different dimensions and for various classes of solutions.

The study by \cite{Hatefi:2021xwh} utilized regression models to estimate nonlinear critical functions. Subsequently, \cite{Hatefi:2022shc} introduced several methods, including truncated power basis, natural spline, and penalized B-spline regression models, to estimate the nonlinear functions relevant to black hole physics, specifically in the axion-dilaton case. Recently, in \cite{Hatefi:2023vma}, artificial neural networks were applied to find black hole solutions within the parabolic class in higher dimensions. Additionally, in \cite{Hatefi:2023sgr}, the Hamiltonian Monte Carlo method was proposed to analyze the complexity of elliptic black holes within a Bayesian framework. Lastly, \cite{Hatefi:2023gpj} employed the sequential Monte Carlo approach to investigate the multimodal posterior distributions of critical functions in hyperbolic equations. This probabilistic approach helps identify existing solutions in the literature and finds all possible solutions that might arise due to measurement errors.

Unlike all existing methods in the literature, we treat the critical exponent in this work as a random variable for the first time, thereby better capturing the uncertainty and numerical measurement errors associated with the perturbed equations of motion. To achieve this, we have developed a novel statistical method that integrates the power and flexibility of Markov Chain Monte Carlo (MCMC) and artificial neural networks, estimating the posterior probability distribution of the critical exponent within a Bayesian framework.
Our approach introduces a generic formalism based on artificial neural networks and the Metropolis-Hastings algorithm \cite{Robert_casella, Murphy} to estimate critical exponent, addressing their inherent complexity. Using quantum perturbation theory, we apply the Bayesian estimation method to determine the critical exponent for the elliptic black hole solution in 4d, estimating the Choptuik exponent within the domain of the equations. We investigate the range of possible values for the critical exponent in the Einstein-axion-dilaton system's 4d elliptic black hole solution.
In this novel iterative approach, each iteration combines the Metropolis-Hastings algorithm with artificial neural networks to simultaneously solve the nonlinear perturbed equations of motion and identify the stochastically most likely values of the critical exponent. We have investigated the Bayesian estimation of the critical exponent based on various perturbed equations independently and by considering all the perturbed equations simultaneously. 
This was done to examine the impact of various perturbed equations on stochastic accept-or-reject transitions. Unlike traditional methods, this new probabilistic approach provides the established solution and explores the range of physically distinguishable critical exponents that may arise due to numerical measurement errors.

This paper is structured as follows: Section \ref{sec:unperturbed} explores the Einstein-axion-dilaton system. Sections \ref{sec:quantum} details the quantum perturbation analysis and the perturbed equations of motion.  Section \ref{sec:stat} presents the statistical methods employed, including polynomial regression, artificial neural networks, and Markov Chain Monte Carlo. Section \ref{sec:num} provides the numerical studies that examine the posterior distribution and Bayesian estimation of the critical exponent.

\section{The Einstein-axion-dilaton System}\label{sec:unperturbed}

The Einstein-axion-dilaton system coupled to gravity in
$d$ dimensions is defined by the following action:
\begin{equation}
S = \int d^d x \sqrt{-g} \left( R - \frac{1}{2} \frac{ \partial_a \tau
\partial^a \bar{\tau}}{(\Im\tau)^2} \right) \; ,
\label{eaction}
\end{equation}
which is known by the effective action of type II string theory \cite{Sen:1994fa,Schwarz:1994xn}. The axion-dilaton is expressed by $\tau \equiv a + i e^{-\phi}$. This action enjoys the $\sltr$ symmetry
 \begin{equation} \label{eq:sltr}
     \tau \rightarrow M \tau \equiv \frac{\alpha \tau + \beta}{\gamma \tau + \delta} \; ,
 \end{equation}
 where $\alpha$, $\beta$, $\gamma$, $\delta$ are real parameters satisfying $\alpha \delta - \beta \gamma = 1$.
If the quantum effects are considered then the $\sltr$ symmetry gets exchanged to $\mathrm{SL}(2,\mathbb{Z})$ and this S-duality was revealed to be the non-perturbative symmetry of IIB string theory~\cite{gsw,JOE,Font:1990gx}. Taking the above action, one derives the equations of motion as
\begin{equation}
\label{eq:efes}
R_{ab} = \tilde{T}_{ab} \equiv \frac{1}{4 (\Im\tau)^2} ( \partial_a \tau \partial_b
\bar{\tau} + \partial_a \bar{\tau} \partial_b \tau)\,,
\end{equation}
\begin{equation}\label{eq:taueom}
\nabla^a \nabla_a \tau +\frac{ i \nabla^a \tau \nabla_a \tau }{
\Im\tau} = 0 \,.
\end{equation}
We take the spherical symmetry for both background and perturbations and the general form of the metric in $d$ dimensions is given by
\begin{equation}
    ds^2 = (1+u(t,r)) (-b(t,r)^2 dt^2 + dr^2) + r^2 d\Omega_{d-2}^2 \,,
\end{equation}
\begin{equation}
    \tau = \tau(t,r).
\end{equation}

The angular part of the metric in $d$ dimensions is $d\Omega_{d-2}^2$.
One can find out a scale-invariant solution by requiring the fact that under a scale transformation  $ (t,r)\rightarrow ( \Lambda t,\Lambda r)$ so that the line element can be changed as $ ds^2 \rightarrow \Lambda^2 ds^2$. Hence all the functions must be scale invariant, i.e. $u(t,r) = u(z)$, $b(t,r) = b(z)$, $z \equiv -r/t$. The effective action~\eqref{eaction} is invariant under the $\sltr$ transformation~\eqref{eq:sltr}, hence
$\tau$ also does need to be invariant up to an $\sltr$ transformation, which means that if the following holds
\begin{equation}\label{eq:tauscaling} 
    \tau(\Lambda t, \Lambda r) = M(\Lambda) \tau(t,r),
\end{equation}
we then call a system of $(g,\tau)$, which respects the above properties of a continuous self-similar (CSS) solution. It is worth highlighting that various cases get related to other classes of $\eval{\dv{M}{\Lambda}}_{\Lambda=1}$~\cite{ours}. Hence,  $\tau$ can take three different forms \cite{AlvarezGaume:2011rk}, called elliptic, hyperbolic and parabolic cases. In this paper, we deal with the elliptic ansatz and its form for the axion-dilaton case can be cast as follows
\begin{align}\label{eq:tauansatz}
\tau(t,r) = i \frac{1-(-t)^{i\omega}f(z)}{1+(-t)^{i\omega} f(z)}\,, & \quad 
\end{align}

where  $\omega$ is a real parameter to be known and the function $f(z)$ needs to satisfy $\abs{f(z)} < 1$ for the elliptic case.
By replacing the CSS ans\"atze in the equations of motion, we would be able to get a differential system of equations for $u(z)$, $b(z)$, $f(z)$. Making use of spherical symmetry, one can show that $u(z),u'(z)$ can be expressed in terms of $b(z)$ and $f(z)$ so that finally, we are left with some ordinary differential equations (ODEs) as follows
\begin{align}\label{eq:unperturbedbp}
    b'(z) & = B(b(z),f(z),f'(z))\,, \\
    f''(z) & = F(b(z),f(z),f'(z))\,. \label{eq:unperturbedfpp}
\end{align}
These equations in elliptic 4d can be written in a closed form as
\begin{eqnarray}
0 & = & b' + { z(b^2 - z^2) \over b (-1 + |f|^2)^2} f' \bar{f}' - {
i \omega (b^2 - z^2) \over b (-1 + |f|^2)^2} (f \bar{f}' - \bar{f} f')
- {\omega^2 z |f|^2 \over b (-1 + |f|^2)^2}, \label{1fzeom320}\nonumber\\
0 & = & f''
     - {z (b^2 + z^2) \over b^2 (-1 + |f|^2)^2} f'^2 \bar{f}'
     + {2 \over (1 - |f|^2)} \left(1
       - {i \omega (b^2 + z^2) \over 2b^2 (1 - |f|^2)} \right) \bar{f} f'^2 \nonumber \\&&
     + {i \omega (b^2 + 2 z^2) \over b^2 (-1 + |f|^2)^2} f f'
\bar{f}' 
  + {2 \over z} \left(1 + {i \omega z^2 (1 + |f|^2) \over (b^2 - z^2)
(1 - |f|^2)}\right.\nonumber \\&& 
+ \left.{\omega^2 z^4 |f|^2 \over b^2 (b^2 - z^2) (1 -
|f|^2)^2}\right) f'+ {\omega^2 z \over b^2 (-1 +|f|^2)^2} f^2
\bar{f}' + \nonumber \\&&
{2i \omega \over (b^2 - z^2)} \left(\frac{1}{2} - {i \omega (1 + |f|^2)
\over 2(1 - |f|^2)}\right.
- \left.{\omega^2 z^2 |f|^2 \over 2b^2 (-1 + |f|^2)^2}
\right) f.
\label{1fzeom321}
\end{eqnarray}
The above equations have five singularities~\cite{AlvarezGaume:2011rk} located at $z = \pm 0$, $z = \infty$ and $z = z_\pm$ where the last singularities are known by $b(z_\pm) = \pm z_\pm$. $z=z_+$ is just a mere coordinate singularity~\cite{Hirschmann_1997,AlvarezGaume:2011rk}, thus $\tau$ should have been regular across it and this constraint actually translates to having the finite value for $f''(z)$ as $z\rightarrow z_+$.

One can actually realize that the vanishing of the divergent part of $f''(z)$ provides us with an equation that is a complex valued constraint at $z_+$ which can be indicated by $G(b(z_+), f(z_+), f'(z_+)) = 0$ where the final form of $G$ 
for the elliptic class is given as \cite{ours}
\begin{eqaed}\label{eq:Gelliptic}
     G(f(z_+),f'(z_+)) = & \, 2 z \bar{f}(z_+) \left(-2 \omega^2\right) f'(z_+)\\ & +f(z_+) \bar{f}(z_+) \left(2 z_+ \bar{f}(z_+) (-2+2 i \omega+2) f'(z_+)+2 i \omega \left(2+\omega^2\right)\right)\\&-\frac{2z_+ (2+2 i \omega-2) f'(z_+)}{f(z_+)}\\&+2 \omega (\omega-i) f(z_+)^2 \bar{f}(z_+)^2-2 \omega (\omega+i)\,.
\end{eqaed}

The initial conditions are known by the smoothness of the solution. By applying polar coordinate as $f(z)=f_m(z) e^{if_a(z)}$, we can see that all equations are invariant under a global redefinition of the phase of $f(z)$, so this means that $f_a(0)=0$. Having set the regularity condition at $z=0$ and making use of the residual symmetries in the equations of motions \eqref{1fzeom321}, we reveal the initial boundary conditions of the equations of motion as follows
\begin{eqaed}\label{bcs}
b(0) = 1, f_m(0) = x_0, \quad\quad\quad\quad   f_m'(0) =f_a'(0)=f_a(0)=0 \;,
\end{eqaed}
where  $x_0$ is a real parameter and $(0<x_0<1)$. Therefore, we have two constraints (the vanishing of the real and imaginary parts of $G$) and two parameters $(\omega,x_0)$. The discrete solution in four dimensions was found in \cite{Antonelli:2019dqv}. Note that these solutions are explicitly found by root-finding procedure as well as numerically integrating all the equations of motion. For instance, the solution in 4d elliptic case was discovered to be \cite{Eardley:1995ns,AlvarezGaume:2011rk} as
\begin{equation}
  \omega=1.176,\quad \abs{f(0)}=0.892,\quad z_+=2.605 \label{esi}
\end{equation}
where the black hole solutions in higher dimensions for the axion-dilaton system are obtained in \cite{Hatefi:2020jdr}. 

 \section{Quantum Perturbation Analysis}\label{sec:quantum}

In this section, we would like to derive the quantum perturbation equations for black holes in the elliptic class in four dimensions. 

This method can be implemented as an extensive method that holds for all dimensions and all matter content.
Some of the steps are taken from ~\cite{Hamade:1995jx} and \cite{Hatefi:2020gis} while we could remove $u(t,r)$ and its derivatives from all the equations, making use of some algebraic computations\footnote{Some similar perturbations of spherically symmetric solutions for Horava Gravity were carried out in ~\cite{Ghodsi_2010}.}. One can perturb the exact solutions $h_0$ (where $h$ denotes either $b$, or $f$) that are explored in Section~\ref{sec:unperturbed} as 
\begin{equation}
 h(t,r) = h_0(z) + \varepsilon \,h_1(t,r)
\end{equation}
where $\varepsilon$ is a small number. By expanding equations in powers of $\varepsilon$, the zeroth order part results in the background equations which have been studied in Section~\ref{sec:unperturbed} and the linearized equations for the perturbations $h_1(t,r)$ are explored due to the linear terms in $\varepsilon $. One can consider the perturbations of the form
\begin{eqnarray}\label{eq:generic_perturbation_ansatz}
h(z,t) = h_{0}(z) + \varepsilon (-t)^{- \kappa} h_{1}(z).
\label{decay}
\end{eqnarray}

The four-dimensional axion-dilaton system is known to be stable, and we consider $\Re \kappa >0$.  We can explore the spectrum of $\kappa$ by solving the equations for $h_1(z)$. Indeed, one can point out that the general solution to the first-order equations will be gained with the linear combination of these modes. We would like to find out the mode $\kappa^*$ with the biggest real part (by assuming a growing mode, which would be $t \rightarrow 0$, i.e. $\Re \kappa > 0$), which will be related to critical Choptuik exponent  as ~\cite{KHA,MA,Hirschmann:1995qx} 
\begin{eqnarray}\label{eq:chopkappa}
\gamma = \frac{1}{\Re \kappa^*}.
\end{eqnarray}
We consider only real modes $\kappa^*$ where the values $\kappa = 0$ and $\kappa = 1$ are indeed gauge modes with respect to phase of $f$ or $U(1)$ re-definitions of $f$ and time translations as well \cite{Antonelli:2019dqv} and these modes have been eliminated from the calculations.

\subsection{Perturbation Equations in 4D For the Elliptic Class}

The derivation of the perturbations has been explained in 
\cite{Antonelli:2019dqv}. However,  we want to analyze the quantum perturbation to explore the entire range of perturbed equations independently and by considering all the perturbed equations simultaneously, as well as their impact on stochastic accept-reject transitions in finding the posterior distributions and Bayesian estimate of the critical exponent. We here highlight briefly all perturbation equations.
The perturbation ansatz~\eqref{eq:generic_perturbation_ansatz}, for  $b$, $\tau$ functions are
\begin{equation}\label{eq:pertb}
    b(t,r) = b_0(z) + \varepsilon\, (-t)^{-\kappa} b_1(z)\ ,
\end{equation}
\begin{equation}\label{eq:pertelliptictau}
    \tau(t,r) = i\frac{1-(-t)^{i\omega}f(t,r)}{1+(-t)^{iw}f(t,r)}\ ,
\end{equation}
\begin{equation}\label{eq:pertf}
    f(t,r) \equiv f_0(z) + \varepsilon (-t)^{-\kappa} f_1(z).
\end{equation}
Using the ans\"atze \eqref{eq:pertb} for the metric functions, we can calculate the Ricci tensor for the metric as a function of $\varepsilon$ to find the zeroth-order and first-order parts from the limiting behaviours as
\begin{equation}
    R^{(0)}_{ab} = \lim_{\varepsilon\rightarrow 0} R_{ab}(\varepsilon)\, \quad  R^{(1)}_{ab} = \lim_{\varepsilon\rightarrow 0} \dv{R_{ab}(\varepsilon)}{\varepsilon}\,.
\end{equation}

The same method is applied to the right-hand side $\tilde{T}_{ab}$ of the field equations, which results in
\begin{equation}
    \tilde{T}^{(0)}_{ab} = \lim_{\varepsilon\rightarrow 0} \tilde{T}_{ab}(\varepsilon)\,\quad    \tilde{T}^{(1)}_{ab} = \lim_{\varepsilon\rightarrow 0} \dv{\tilde{T}_{ab}(\varepsilon)}{\varepsilon}.
\end{equation}
Indeed, the Einstein Field Equations (EFEs) are held order by order so that
\begin{equation}
    R^{(0)}_{ab} = \tilde{T}^{(0)}_{ab}\,,\quad R^{(1)}_{ab} = \tilde{T}^{(1)}_{ab}.
\end{equation}
As explained, one can remove $u$  and its first derivative in terms of $b_0,f$ and their first derivatives. We now combine the field equations so that we would be able to eliminate the second-derivative terms in $b(t,r)$. This procedure is known as Hamiltonian constraint, which can be shown by
\begin{equation}
    C(\varepsilon) \equiv R_{tt} + b^2\,R_{rr} - \tilde{T}_{tt} - b^2\, \tilde{T}_{rr} = 0\,.
\end{equation}
One finds the lowest-order contribution as a first-order equation relating $b_0'$ to $b_0$, $f_0$, $f_0'$ as follows
\begin{eqnarray}\label{eq:ellipticb0p}
b_0'= \frac{((z^2-b_0^2)f_0'(z \bar f_0'+i \omega  \bar f_0)+\omega f_0(\omega z \bar f_0-i(z^2-b_0^2) \bar f_0'))}{b_0 (1-f_0 \bar f_0)^2}\,.
\end{eqnarray}
In a similar way, the first correction is given by 
\begin{equation}
    \eval{\dv{C(\varepsilon)}{\varepsilon}}_{\varepsilon=0} =0 \ ,
\end{equation}
which is a first-order equation relating $b_1'$ to $b_0$, $b_0'$ $f_0$, $f_0'$, $f_0''$, and the other perturbations $b_1$, $f_1$, $f_1'$, which are indeed linear in all perturbations. For the elliptic class in 4d one obtains the linearized equations for $b_1'$ as  follows 
\begin{eqaed}\label{eq:equationb1p_elliptic}
(L_1)b_1'&=r t ((t-2 t)b_0+rb_0')\bigg(-\frac{2 f_1' \bar f_0' r^2}{t^4 s_0^2}-\frac{2 b_1 b_0'}{r t}+\frac{\kappa 2 b_0 b_1 b_0'}{r \left((t-2 t) b_0+rb_0'\right)}\\& \quad
-\frac{4 i \omega b_0 b_1 \bar f_0 f_0'}{r t s_0^2}
+\frac{2i \omega b_0^2 \bar f_1 f_0'}{r t s_0^3}
+\frac{2\kappa \bar f_1 f_0'r}{t^3 s_0^2}+\frac{2i  \omega \bar f_1 f_0' r}{t^3 s_0^2}
\\& \quad 
-\frac{2 \kappa b_0^2 \bar f_1 f_0'}{r t s_0^2}
+\frac{2i \omega \bar f_0  f_1'r}{t^3 s_0^2}
-\frac{2 i \omega b_0^2 \bar f_0 f_1'}{r t s_0^2}
+\frac{4 b_0 b_1 f_0' \bar f_0'}{t^2 s_0^2}
 \\& \quad 
+\frac{2b_0^2 f_1' \bar f_0'}{t^2 s_0^2}
+\frac{1}{r t^4 s_0^3} 2f_1\Big(t \omega \bar f_0^2 (rt(-i\kappa+\omega)f_0-2i(r^2-t^2b_0^2)f_0')
\\& \quad \quad
+t(\kappa-i\omega)(-r^2+t^2b_0^2) \bar f_0'
+\bar f_0 \big(rt^2 \omega(i\kappa+\omega)+t(\kappa+i\omega)
\\& \quad \quad \times(r^2-t^2b_0^2)f_0\bar f_0'
+2r (r^2-t^2b_0^2)f_0'\bar f_0'\big)\Big)
-\frac{2 f_0' \bar f_1' r^2}{t^4 s_0^2}
+\frac{2 b_0^2 f_0' \bar f_1'}{t^2s_0^2} \\
& \quad -\frac{2if_0 (\bar f_1(r\omega t^2 (\kappa+i\omega)+\omega t(2r^2-t^2 b_0^2)\bar f_0 f_0'+2ir(r^2-t^2 b_0^2) f_0' \bar f_0'))}{rt^4s_0^3}
\\& \quad
-\frac{2 i f_0 t \omega(-\bar f_1' r^2+2t^2 b_0 b_1 \bar f_0'+t^2 b_0^2 \bar f_1')}{rt^4s_0^3}
+\frac{4 i \omega f_0^2 \left(r^2-t^2 b_0^2\right) \bar f_1 \bar f_0'}{r t^3 s_0^3}
\\&
+\frac{2 i \omega f_0^2 \left(\bar f_0 \left(-\bar f_1' r^2+t (\kappa-i \omega) \bar f_1 r+2 t^2 b_0 b_1 \bar f_0'+t^2b_0^2 \bar f_1'\right)\right)}{r t^3 s_0^3}\bigg)\,,
\end{eqaed}
where
\begin{equation}
    L_1 = 2 b_0 ((\kappa-1) tb_0+rb_0'), \quad  s_0 =(f_0 \bar{f}_0-1).
\end{equation} 
It can be seen that \eqref{eq:equationb1p_elliptic} is invariant under dilations $(t,r)\rightarrow (e^\lambda r,e^\lambda t)$, so that, by changing coordinates to $(t,z)$, one finds that all the factors of $t$ will cancel off and the final result just depends on $z$. Hence, we are left with just real and linear ordinary differential equations. 

If we apply the perturbation ans\"atze in the $\tau$ equation of motion~\eqref{eq:taueom} and extract the zeroth and first-order parts and remove all $u_0'$, $u_1$, $u_1'$, then the resulting zeroth-order part will be an equation including $b_0$, $b_0'$ $f_0$, $f_0'$, $f_0''$, while substituting $b_0'$ according to \eqref{eq:unperturbedbp} and solving for $f_0''$, we obtain a second-order equation for $f_0$, which is indeed the explicit form of \eqref{eq:unperturbedfpp}.
The first-order part includes $b_1$, $b_1'$, $f_1$, $f_1'$, and $f_1''$ linearly, while the equations depend on the zeroth-order functions as well as their derivatives non-linearly as background functions. Hence, we are left with the perturbations for $f_1''$ given by
\begin{eqaed}
  (L_3)f_1''=&r^2\bigg( it \omega f_0^2(\bar f_1b_0'+\bar f_0 b_1') +f_1b_0'(\kappa t-it \omega -t(\kappa-2i\omega)f_0\bar f_0+r f_0'\bar f_0)
  \\
  &\quad -r( b_1'f_0'+b_0'f_1')+f_0\big(b_1'(-it \omega+r\bar f_0 f_0')+rb_0'(\bar f_1 f_0'+\bar f_0 f_1')\big)\bigg)\\
  & +rt^2b_0^2\big(f_1\bar f_0 b_0'  f_0'-b_1' f_0'-b_0' f_1'+f_0(\bar f_1 b_0' f_0'+\bar f_0(b_1'f_0'+b_0'f_1'))\big)\\
  & -t^2b_0^3\bigg(2r\bar f_1f_0'^2-2tf_1'+4r\bar f_0 f_0' f_1'+f_1 \bar f_0(2tf_0'-rf_0'')\\
  & \quad  +f_0\big(2t\bar f_0 f_1'+\bar f_1(2tf_0'-rf_0'')\big)\bigg)\\
  &-rb_0\bigg(t^2\omega(\omega-i)f_0^2\bar f_1 +2r\big(-r \bar f_1 f_0'^2+(t(1+\kappa-i\omega)-2r\bar f_0 f_0')f_1'\big)\\
  & \quad \quad  +f_1\Big(t^2(-\kappa^2+\kappa(-1+2i \omega)+ \omega (i+\omega))+t^2(\kappa+\kappa^2+2i\kappa\omega \\
  &  \quad \quad \quad +2\omega(-i+\omega)\big)\bar f_0 f_0+r\bar f_0(2t(-1+2\kappa-i\omega)f_0'+rf_0'')\Big)\\
  & \quad \quad +rf_0(-2t(1+\kappa+i\omega)\bar f_0 f_1'+\bar f_1(-2it(-i+\omega)f_0'+rf_0''))\bigg) \\
  & -b_1\bigg(rt^2 \omega(-i+i\kappa+\omega)f_0^2\bar f_0-t\big(r^2(-2+\kappa+2i\omega)+6t^2b_0^2\\
  &\quad -2rtb_0b_0'\big)f_0'
  -2r(r^2-3t^2b_0^2)\bar f_0f_0'^2-r(r^2-3t^2b_0^2)f_0''
  \\
  &\quad +f_0\Big(rt^2\omega(i-i\kappa+\omega)+
  \bar f_0\big(t(r^2(-2+\kappa-2i\omega)+6t^2b_0^2\\
  &\quad -2rtb_0b_0')f_0'
  +r(r^2-3t^2b_0^2)f_0''\big)\Big)\bigg)\,,
  \end{eqaed}
where
\begin{equation}
    L_3= r b_0(r^2-t^2b_0^2) s_0\,.
\end{equation}
The above equations are scale-invariant. Hence, they turn to an ordinary differential equation after making use of a change of coordinates to $(z,t)$. Thus, the system of ordinary linear differential equations is given by
\begin{align}
    b_1' & = B_1(b_1,f_1,f_1')\,,\label{eq:equationb1p}\\
    f_1'' & = F_1(b_1,f_1,f_1')\,.\label{eq:equationf1pp}
\end{align}
Note that $b_1$  and $f_1$ are now linear functions that still depend non-linearly on the unperturbed solution. We also have  a quadratic dependence on $\kappa$ as well. Note that these equations also do include the same singularities as appeared in the unperturbed system of equations which means that they are also singular for $z=0$ and $b^2(z)=z^2$. The modes are now explored by finding the $\kappa$ values that are related to smooth solutions of the perturbed equations~\eqref{eq:equationb1p},~\eqref{eq:equationf1pp}  where they need to satisfy the proper boundary conditions as follows.

We now focus on the boundary conditions needed to solve
\eqref{eq:equationb1p} and \eqref{eq:equationf1pp}. First of all, at $z=0$, we re-scale the time coordinate so that $ b_1(0) = 0$. Also, using the regularity condition for the axion-dilaton at $z=0$, we find that $ f_1'(0) = 0$ so that the freedom in $f$ is reduced to $f_1(0)$ which is an unknown complex parameter.  On the other hand, at  $z_+$ (we recall that $z_+$ is defined by the equation $b(z_+)=z_+$) all equations and perturbations are regular
so that all the second derivatives $\partial_r^2 f(t,r)$, $\partial_r \partial_t f(t,r)$, $\partial_t^2 f(t,r)$ should be finite as $z\rightarrow z_+$. Hence, $f_0''(z)$ and $f_1''(z)$ are also finite as $z \rightarrow z_+$. We introduce $\beta = b_0(z)-z$ and then expand $f_0'', f_1''$ near the singularity, as
\begin{align}\label{eq:taylorexpansion_unperturbed}
    f_0''(\beta) & = \frac{1}{\beta} G(h_0) + \mathcal{O}(1)\,,\\ \label{eq:taylorexpansion_linearised}
    f_1''(\beta) & = \frac{1}{\beta^2} \bar{G}(h_0) + \frac{1}{\beta} H(h_0, h_1|\kappa) + \mathcal{O}(1)\,,
\end{align}
where we have defined the following equations:
\begin{align}\label{eq:taylorexpansion_unperturbed}
h_0 = (b_0(z_+),f_0(z_+),f_0'(z_+)),\quad h_1 = (b_1(z_+),f_1(z_+),f_1'(z_+))\;.
\end{align} 
The vanishing unperturbed complex constraint is given by $G(h_0)=0$ at $z_+$, and  we checked that it implies $ \bar G(h_0) = 0$  at $z_+$ as well, which means that
\begin{equation}
    G(h_0) = 0 \,\Rightarrow \, \bar{G}(h_0) = 0\,.
\end{equation}

Hence, we are left just with the complex-valued constraint 
\begin{equation}\label{eq:constraintH}
    H(h_0, h_1 | \kappa) = 0\,,
\end{equation}
that is linear in $h_1$. 

We now try to solve this constraint for $f_1'(z_+)$ in terms of $f_1(z_+)$, $b_1(z_+)$, $\kappa$ and $h_0$. Hence,  this condition reduces the number of free parameters in the boundary conditions at $z_+$ to a real number $b_1(z_+)$ and a complex $f_1(z_+)$. In conclusion, we have six real unknowns, which are $\kappa$ and the five-component vector as
\begin{equation}\label{eq:defX}
    X = (\Re f_1(0),\, \Im f_1(0),\, \Re f_1(z_+),\, \Im f_1(z_+),\, b_1(z_+) )\;,
\end{equation}
and the system of linear ODE's \eqref{eq:equationb1p} and \eqref{eq:equationf1pp} where the total real order is indeed five.
From the numerical procedure of \cite{Antonelli:2019dqv} and given a set of boundary conditions $X$, we integrate from $z=0$ to an intermediate point $z_\text{mid}$, and similarly, we integrate backwards from $z_+$ to $z_\text{mid}$. Finally, we collect the values of all functions $(b_1, \Re f_1, \Im f_1, \Re f_1', \Im f_1')$ at $z_\text{mid}$ and encode the difference between the two integrations in a difference function $D(\kappa;X)$. By definition, $D(\kappa; X)$ is linear in $X$ thus it has a representation in matrix form
\begin{equation}
    D(\kappa; X) = A(\kappa) X \;,
\end{equation}
where $A(\kappa)$ is a $5\times 5$ real matrix depending on $\kappa$. Thus, we only need to find the zeroes of $D(\kappa; X)$ and this can be achieved by evaluating $\det A(\kappa) = 0$. We carry out the root search for the determinant as a function of $\kappa$ where the root with the biggest value will be related to  the Choptuik exponent through \eqref{eq:chopkappa}. Note that the perturbed equations of motion are singular whenever the factor $\left(\kappa-1-z \frac{b_0'}{b_0}\right)$ in the denominator vanishes, so that the numerical procedure fails at a particular point. We can get an estimate for the values of $\kappa$ giving rise to this singular behaviour, and we can also find the entire region that leads to numerical failure. However, this apparent problem does not affect our evaluation of the critical exponent because, in most cases, the most relevant mode $\kappa^*$ lies outside that particular failure region.

\section{Statistical Approach}\label{sec:stat}

\begin{figure}
\centering
\includegraphics[width=1\textwidth]{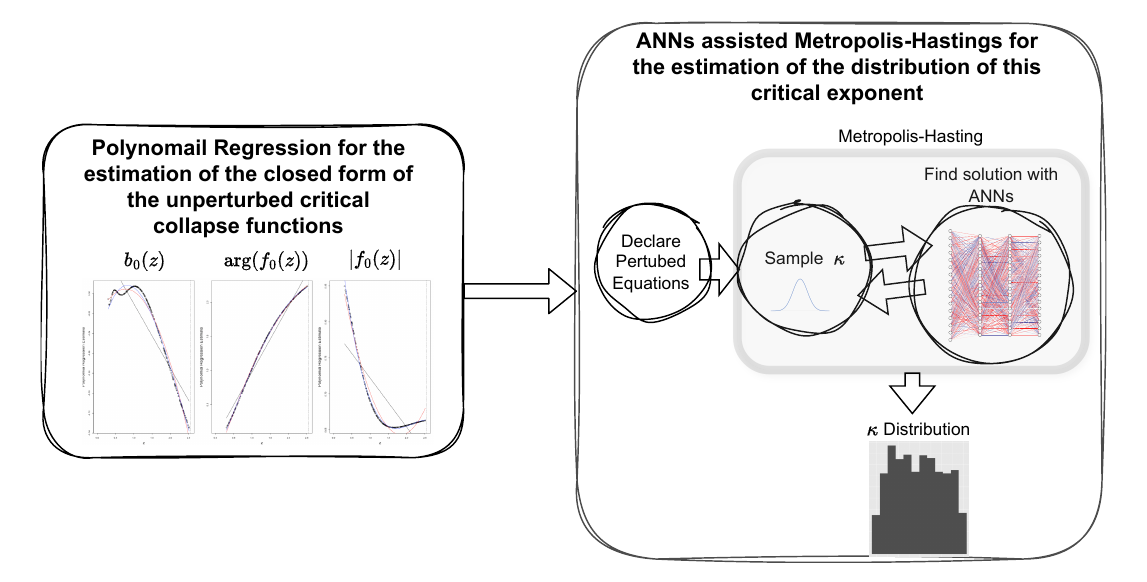}
\caption{Bayesian approach with artificial neural networks assisted Metropolis-Hastings for the estimation of the distribution of $\kappa$. We start with the estimation of the closed form for the unperturbed equations of motion following polynomial regression. Then, a Bayesian artificial neural networks assisted Metropolis-Hastings is followed, sampling $\kappa$ from a prior distribution, and using an ANNs solver for the solution of the perturbed equations of motion in each of the Metropolis-Hastings iterations.}\label{fig:pipeline}
\end{figure}

In this section, we detail the novel approach we have developed to estimate the distribution of the critical exponent $\kappa$ in the elliptic 4d case of the equations of motion with perturbations.
In figure~\ref{fig:pipeline} we show the main steps of the proposed pipeline.

Technically, we offer a new Bayesian approach with artificial neural networks (ANNs) assisted Metropolis-Hastings for the estimation of the distribution of the mentioned $\kappa$ that is related to the critical exponent $\gamma$. 
To do so, we must treat $\kappa$ a random variable, so we use the Bayesian strategy to find the posterior distribution of $\kappa$ based on the perturbed DE system. 
We emphasize again that this is first time in the literature where the critical exponent is treated as a random variable.
As can be seen in figure~\ref{fig:pipeline}, our approach starts with the use a polynomial regression technique to estimate the closed form of the unperturbed critical collapse functions under analysis.
To properly feed the polynomial regression technique, we follow an ANN-based approach~\cite{Hatefi:2023sgr} to numerically find the unperturbed critical collapse functions in the entire domain of the DE system.
Our approach uses the ANN estimates of the unperturbed critical collapse functions so that the polynomial regression technique can be used for the estimation of the final closed form.
As our objective is the Bayesian analysis of the perturbed equations, we use the estimated closed-form to incorporate the perturbations related to $\kappa$.
The last step consists of the implementation of ANN-assisted Metropolis-Hastings to estimate the distribution of of $\kappa$ which is related to the critical exponent.
We detail each step in the following subsections, providing all the details for the statistical methods employed.

\subsection{Polynomial regression for the estimation of the closed form of the unperturbed equations}\label{sub:poly}
Polynomial regression models are common tools in data science to model multivariate data and find the relationship between variables. 
Let $({\bf x},{\bf Z})$ be the input multivariate data, where ${\bf x}$ denotes the response vector of sample size $n$ and ${\bf Z}$ denotes the design matrix of size $(n \times p)$ with $p$ independent variables $({\bf z}_1,\ldots,{\bf z}_p)^\top$. Given the training data of size $n$, the multivariate regression 
\begin{align}\label{reg}
    x_i = {\bf z}_i^\top {\bf \beta} + \epsilon_i,
\end{align}
where $\epsilon_i$ are independent and identically distributed (iid) variables from the standard Gaussian distribution. 
The regression model \eqref{reg} essentially converts predicting the response differential equation variable $x=g({\bf z})$ into estimating the unknown coefficients of the model. 
The least squares (LS) method is a common approach to estimating the regression model coefficients \eqref{reg} minimizing the $l_2$ norm between the predicted responses and observed responses as
\begin{align}\label{l2_norm}
{\widehat \beta} = \arg \min_\beta || {\bf x} - {\bf Z} {\bf \beta} ||^2_2. 
\end{align}
One can easily show that the LS estimate of unknown coefficients can be obtained by $\widehat{\bf \beta}_{LS} = ({\bf Z}^\top {\bf Z})^{-1} {\bf Z} {\bf x}$ as the solution to \eqref{l2_norm}. 
Hence, the differential equation response $x(z)$ at every space-time point
$z_{new}$ is predicted by ${\widehat x}_{new} = {\bf z}_{new} {\widehat\beta}_{LS}$ \cite{poly_harrel}. 

The polynomial regression model extends the properties of multivariate regression by employing the higher-order terms of the independent variables to better predict the nonlinear pattern of the response function $x=g(z)$. The polynomial regression of order $J$ is given by 
\begin{align}\label{pol_reg}
    x_i = \sum_{j=l}^{ J} z_i^j \beta_j + \epsilon_i, 
\end{align}
where $\beta= (\beta_1,\ldots,\beta_J)$ are the model coefficients that should be estimated. 
One can easily represent \eqref{pol_reg} in matrix form as \eqref{reg} where the columns of the design matrix  ${\bf Z}$ induce the explanatory variable $z_i,i=1,\ldots,n$ raised to various polynomial powers $j=1,\ldots,J$. 
From the least squares method \eqref{l2_norm}, one can predict the non-linear responses of the differential equations at every point $z_{new}$ in the domain of the equations of motion by 
${\widehat x}_{new} = {\bf z}_{new} {\widehat\beta}_{LS}$.

Technically, we follow the described polynomial regression technique to first obtain a closed form of the unperturbed equations of motions.
As it has been detailed, the polynomial regression needs pairs of corresponding inputs and outputs.
For generating them, for our problem, we follow an ANNs-based approach~\cite{Hatefi:2023sgr} to numerically find the unperturbed critical collapse functions in the entire domain of the DE system.
In other words, we employ the technique in~\cite{Hatefi:2023sgr} to generate the training data of the polynomial regressor.

\subsection{ANNs assisted Metropolis-Hastings for the estimation of the distribution of the critical exponent}\label{sub:ann_mcmc}

We do have a closed form of the unperturbed equations of motion thanks to the polynomial regression detailed in previous section.
However, our objective consists of finding the distribution of the parameter of $\kappa$ in the perturbed equations of motion.

With the estimations of the closed form for the DE variables $b_0(z),|f_0(z)|$ and $\arg(f_0(z))$ of the unperturbed equations of motion, we proceed to incorporate these estimates and update the perturbed equations of motion \eqref{eq:equationb1p} and \eqref{eq:equationf1pp}. We consider these perturbed equations of motion, as our underlying DE system to estimate the distribution of the critical exponent.

Consider a system of $H$ differential equations where ${\bf x}(t) = \left(x_1(t),\ldots, x_H(t)\right)$ represents the multivariate solutions evaluated at space-time $t$ to the system of differential equations (DEs)
\begin{align}
    \frac{d}{dt} x_h(t) = g_h({\bf x}(t)|\theta),
\end{align}
where the DE variable $x_h(t)$ corresponds to $h$ DE and $\theta$ denotes the collection of the unknown parameters of the DE. 
In our particular case $\theta = \kappa$.
Due to the high non-linearity of the black hole equations of motion, the exact solution to the DE equations can be observed. 
Rather than the exact DE solution, one may observe a perturbed version of the DE variable with numerical measurement errors through numerical experiments. 
To take this uncertainty into statistical methods, let $ {\bf y}_h = ( y_{h1},\ldots,y_{hn})$ denote the observed version of the DE variable $x_h(t)$ at $n$ space-time points. Hence, let $y_{h,i}$ follow a Gaussian distribution with mean $x_h(t_i|\theta)$ and variance $\sigma_h^2$. Given the Gaussian distribution of the observed data, the uncertainty in the DE data can be modelled by the likelihood function of $y_i$ as
\begin{align}\label{like}
    L(\theta|{\bf y}) = \prod_{h=1}^{H} \prod_{i=1}^{n} (1/\sigma_i^2) 
    \exp\left\{ - {(y_{hi} - x_h(t_i|\theta)}/{2 \sigma_h^2}\right\},
\end{align}
The likelihood function translates the problem of finding the solution to a system of DEs into a Gaussian process with unknown parameters. 
One can estimate the unknown parameters of the DE system given the observed DE variables by maximizing the likelihood function. 

According to the non-linear nature of the DE variables, the likelihood function \eqref{like} appears nonlinear with multiple optimums, indicating the sensitivity of the maximum likelihood estimate of the DE variables.  
Technically, in this work, we propose a Bayesian approach incorporating the numerical measurement errors in estimating the critical exponent parameter in the equations of motion \eqref{eq:equationb1p},\eqref{eq:equationf1pp}. This framework requires prior knowledge of model parameters. 
This prior knowledge is incorporated into equations through a prior probability density function. 
Let $\pi(\theta;{\bf \alpha})$ denote this prior distribution, with ${\bf \alpha}$ representing the vector of hyper-parameters. 
In the experiments, see Section~\ref{sec:num}, we employ different prior distributions.

So, we are able to first generate $\kappa$ candidates from a proposal distribution. 
Then, we propose to apply fully connected ANNs using the $\kappa$ candidate and find the solution to the perturbed equations of motion corresponding to the $\kappa$ candidate.
This means that our model will generate for each $\kappa$ candidate a solution using ANNs, assisting in an iterative fashion the deployed Bayesian approach.
We show now in detail how this ANNs step is done.

Standard ANNs encompass multi-layer perceptrons, which transfer the data between the layers through linear and non-linear functions \cite{Goodfellow:Deep}.
In this research, we focus on fully connected ANNs to deal with the non-linearity of the elliptic perturbed equations of motion in the Einstein-axion-dilaton system in 4d. 
Let $\mathcal{N}^{L}({\bf x}, t, \psi)$ denote a neural networks with $L$ layers which map the input dimensions $R^{in}$ to  $R^{out}$. 
Let ${\bf W}^l$ and ${\bf b}^l$ denote the weight matrix and bias vector, which regress the neurons in layer $l$ on $l-1$, respectively. 
Accordingly, the response vector before activation in layer $l$ is given by 
\begin{align*}
\begin{array}{ll}
    \text{input layer:} &   N^{0}({\bf x},t,\psi) = {\bf x},\\
    \text{hidden layers:} & N^{l}({\bf x},t,\psi) = \eta({\bf W}^l N^{l-1}({\bf x},t,\psi) + {\bf b}^l),\\
    \forall 1 \le l\le L-1 & \\
    \text{output layer:} & N^{L}({\bf x},t,\psi) = \eta({\bf W}^L N^{L-1}({\bf x},t,\psi) + {\bf b}^L),\\
\end{array}    
\end{align*}
where the output of layer $l$ is obtained after applying the activation function, that is $\eta(\cdot)$.

The universal approximation theorem guarantees that the neural networks can approximate any function, making the neural networks a state-of-the-art method for solving nonlinear equations \cite{Goodfellow:Deep}. One must define a loss function $\mathcal{L}$ to assess the performance of the ANNs and a back-propagation algorithm to adjust the weight matrices and bias parameters of the ANNs based on the gradient of the loss functions. The back-propagation algorithm calculates the gradients of the ANNs with respect to the response variable to ensure that the estimates of the ANN's coefficients meet the requirements of the loss function. 
This process translates solving the differential equations of motion into an optimization problem, estimating the differential variables based on minimizing the squared residues of the ANNs.  
Accordingly, for this purpose, we introduce the $l_2$ loss function as 
\[
\mathcal{L}\left(\mathcal{N}^{L}({\bf x}, t, \psi)\right)= \left(\mathcal{N}^{L}({\bf x}, t, \psi) - g \right)^2,
\]
 where $\mathcal{N}^{L}({\bf x}, t, \psi)$ represents the trained neural networks approximating the underlying function. 

Overall, at this point, our method disposes of a solution of the system of DEs pertinent to the perturbed elliptic class of black hole equations of motion, for a sampled particular value of $\kappa$.
However, our objective is to find the probability distribution of the critical exponent, so that it can be considered, for the first time in the literature, as a random variable.
Therefore, for each of the solutions associated to sampled values of $\kappa$ we must compute the likelihood function \eqref{like} and find its acceptance probability 

Using information contained in the likelihood function \eqref{like} and the prior distribution, one can find the posterior distribution of the unknown parameters $\pi({\bf\theta}|{\bf y})$. The posterior distribution enables us to capture the pattern of the parameters of the DE system using all the uncertainty involved in calculating the DE variables. The posterior distribution of $\theta$ is calculated by 
\begin{align}\label{post_dis}
    \pi({\bf\theta}| {\bf y}) = \frac{L({\bf\theta}|{\bf y}) \pi({\bf\theta};{\bf \alpha})}{ \int_\theta L({\bf\theta}|{\bf y}) \pi({\bf\theta};{\bf \alpha})}. 
\end{align}
Nevertheless, the posterior distribution \eqref{post_dis} lacks a closed-form solution because of the high non-linearity of the DE variables and the complex marginal distributions of the DE variables. For this reason, one has to use iterative methods such as Markov Chain Monte Carlo to find numerically the posterior distribution of the parameters of the DE system \cite{Girolami_2008}.  

Markov Chain Monte Carlo (MCMC) is a widely acceptable numerical technique within the Bayesian framework to estimate the posterior distribution of the DE system. The power of the MCMC approach relies on the fact that it transfers the estimation task into the sampling process from the posterior distribution of interest.
Extensive research has been conducted in the literature about the properties of the MCMC approach, including Metropolis-Hastings, Gibbs sampling, and Sequential Monte Carlo \cite{bishop} and \cite{Murphy}. In its nutshell,  MCMC employs various probabilistic techniques to generate a Markov chain of samples approximating the target posterior distribution.  
When the chain is selected long enough, the Markov chain will eventually reach a stationary state that accurately fluctuates around the true form of the posterior distribution. 

For our implementation, we follow the Metropolis-Hastings (MH) algorithm \cite{Robert_casella}, which is an established MCMC approach. 
The MH method calculates the probability of the transition between the current state of the chain $\theta^{(t)}$ and a candidate state $\theta^*$ simulated stochastically from an independent proposal distribution $q(\theta)$. The MH method applies a probabilistic step to accept or reject the proposed candidate.  
Let $\theta^{(t)}$ represent the $t$-th state of the MCMC chain. Then we accept the proposed state $\theta^*$ as the next state of the chain with probability 
\begin{align}\label{p_mh}
    \min \left\{1, \frac{\pi({\bf\theta}^* | {\bf y}) q({\bf\theta}^{(t)}) }{\pi({\bf\theta}^{(t)}| {\bf y}) q({\bf\theta}^*)} \right\}.
\end{align}
If the candidate ${\bf\theta}^*$ is accepted, then ${\bf\theta}^{(t+1)}={\bf\theta}^*$; otherwise, ${\bf\theta}^{(t+1)}={\bf\theta}^{(t)}$. 
The entire MCMC method is replicated $N$ times to generate a sequence of $\{{\bf\theta}^{(t)}; t=1,\ldots,N\}$ samples from the target posterior distribution $\pi({\bf\theta} | {\bf y})$. 

Overall, we have detailed a novel model that allows for a Bayesian estimation of the critical exponent on the elliptic black hole solution in 4d.
In the next section, we show all the numerical studies performed, and for the first time in the literature, we are able to provide a probability distribution of $\kappa$.

\section{Numerical Studies}\label{sec:num}
In this section, we plan to find the distribution of the critical exponent $\kappa$ in the elliptic 4d equations of motion in the Bayesian framework. To do so, the critical exponent is treated as a random variable, so we use the Bayesian strategy to find the posterior distribution of $\kappa$ based on the perturbed DE system. In this numerical study, we investigate the equations of motion as the input DE system. 

Hatefi et al. (2023) in \cite{Hatefi:2023sgr} proposed ANNs to solve the unperturbed DE variable.  We applied the properties of polynomial regression and found the closed-form polynomial regression estimates for the unperturbed critical collapse functions. 
According to the fact that the goal of this research is to estimate the critical exponent and that the critical exponent is only observed in perturbed DE variables, we first followed \cite{Hatefi:2023sgr} and used the ANNs based on $\omega=1.176$ to numerically find the unperturbed critical collapse functions in the entire domain of the DE system. 
To find the regression estimate of the unperturbed critical collapse functions, we generated equally spaced points  ${z_1,\ldots,z_n}$ of size $n=1000$ from the domain of the DE system. 
We obtained the ANN estimates of the unperturbed critical collapse functions evaluated at each of the 1000 points.   
Once we obtained the ANN estimates, we used 750 observations as training and 250 as testing data to estimate the closed-form polynomial regression of the unperturbed critical collapse functions, as described in Subsection \ref{sub:poly}. 
Hence, the closed-form polynomial estimates of the unperturbed critical collapse functions are derived by 
\begin{align}\label{b0_est}
    \widehat{b_0(z)} &=  1.005 - 0.187 z + 0.480 z^2 - 0.004 z^3,  \\\label{b0_est}
    \widehat{|f_0(z)|} &=  0.919 - 0.122 z - 0.028 z^2 - 0.002 z^3,  \\\label{b0_est}
    \widehat{\arg(f_0(z))} &=  -0.011 + 0.041 z + 0.047 z^2 - 0.012 z^3.
\end{align}

\begin{figure}
\centering
\includegraphics[width=1\textwidth]{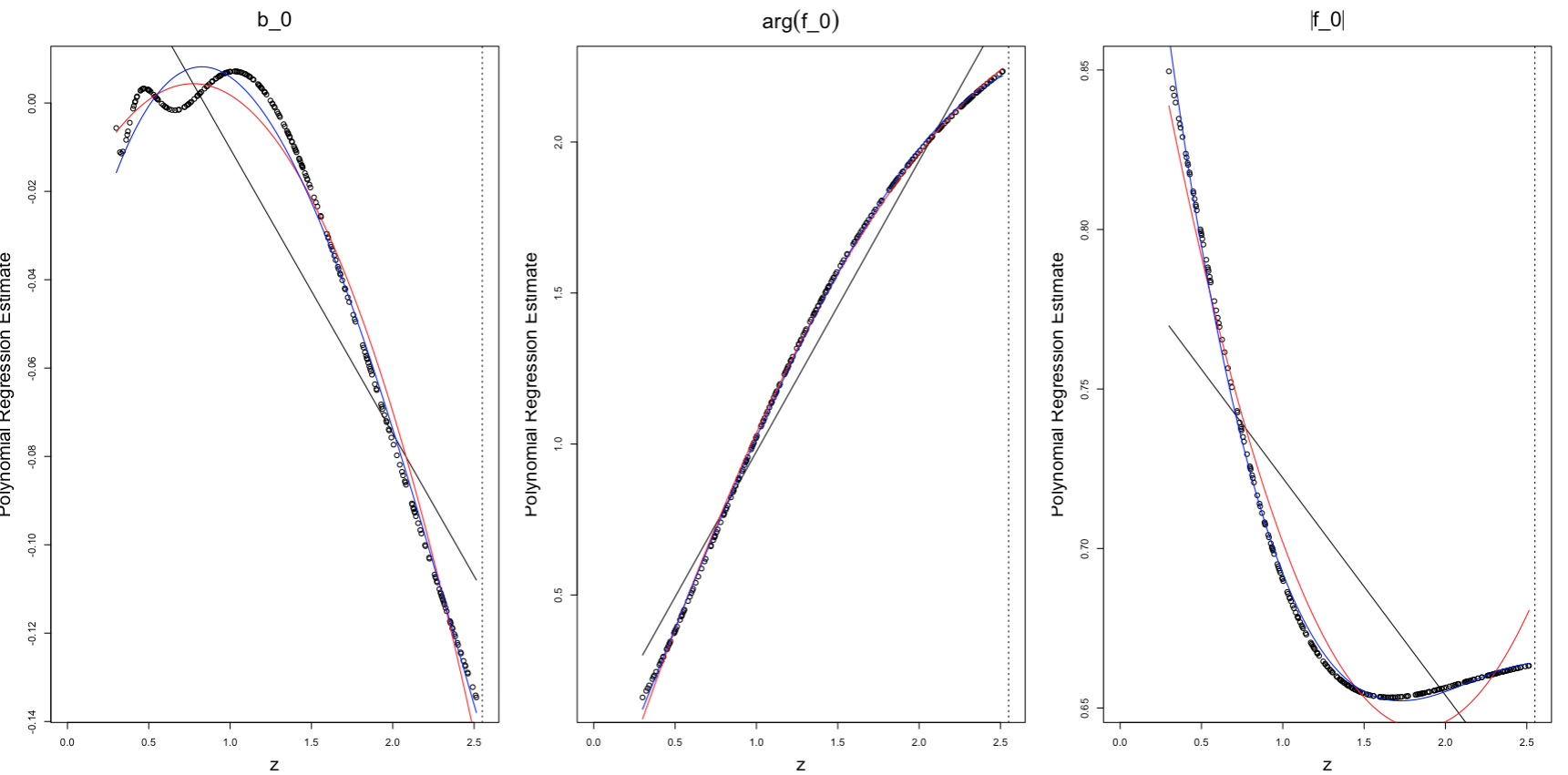}
\caption{The polynomial estimates of the unperturbed critical collapse functions  $b_0(z), \arg(f_0(z))$ and $|f_0(z)|$ of the equations of motion based on polynomial orders of $l=1$ (black), $l=2$ (red) and $l=3$ (blue). The dots show the true functional form evaluated at the validation points.}   \label{fig_poly_unp}
\end{figure}

Figure \ref{fig_poly_unp} shows the training and test performance of the polynomial estimates for the critical collapse functions for polynomial orders $l=1,2,3$. One can easily see that the polynomial of order $l=3$ can capture very well the non-linearity of the unperturbed critical collapse functions. 
Once we found the unperturbed DE variables $b_0(z),|f_0(z)|$ and $\arg(f_0(z))$, we incorporated these estimates and updated the perturbed equations of motion. In the next step, we consider this perturbed equations of motion, as our underlying DE system to estimate the critical exponent.

We proposed an artificial neural network-assisted Metropolis-Hastings to obtain the posterior distribution of $\kappa$. We first generated $\kappa$ candidates from a proposal distribution to do so. 
Here, we considered two proposal distributions, including the Uniform distribution between (3,4) and the Gaussian distribution with a mean of 3.5 and variance of 1 to deal with the singularity issue, and the range included the established solution available in the literature \cite{Antonelli:2019dqv}.

In the next step, we apply fully connected neural networks using the $\kappa$ candidate and find the solution to the perturbed equations of motion corresponding to the  $\kappa$ candidate. We implemented the fully connected  ANNs using the Python package neurodiffeq  \cite{Chen:2020}.
The neural networks used contain 4 hidden layers, each with 16 neurons.
The training was done for 50 epochs.  
We then computed the likelihood function \eqref{like} and found the acceptance probability of the Metropolis-Hasting algorithm under non-informative uniform distribution. 
Due to the fact that the range of DE variables is dramatically different, we assign $\sigma_{b_1} = 7$ and $\sigma_g=2$ in the likelihood function of $\kappa$ given the perturbed DE variables. 

It is shown in the literature that the perturbed $b_1(z)$ DE variable is linear with respect to $\kappa$; however,  the perturbed $f_{1m}(z)=g_m(z)$ contains the higher orders of $\kappa$ \cite{Antonelli:2019dqv}. 
To investigate the effects of the different DE variables on the MCMC chains and the Bayesian distribution of $\kappa$, we evaluated the performance of the stochastic accept-reject method based on three scenarios. 
These scenarios include the accept-reject step based on only the observed perturbed DE variable of $b_1(z)$, accept-reject based on only the observed 
perturbed DE variable $g_m(z)$ and accept-reject based on both observed perturbed DE variables $g_m(z)$ and $b_1(z)$. 
Finally, the entire neural network-assisted Metropolis-Hasting approach was carried out for 2000 chains, where in each iteration, the perturbed equations of motion were computed using artificial neural networks. 
We discard the first 10\% of the history of the MCMC chain to wash out the effect of the initialization step on the performance of the MCMC chain.  

\begin{figure}
\centering
\includegraphics[width=1\textwidth]{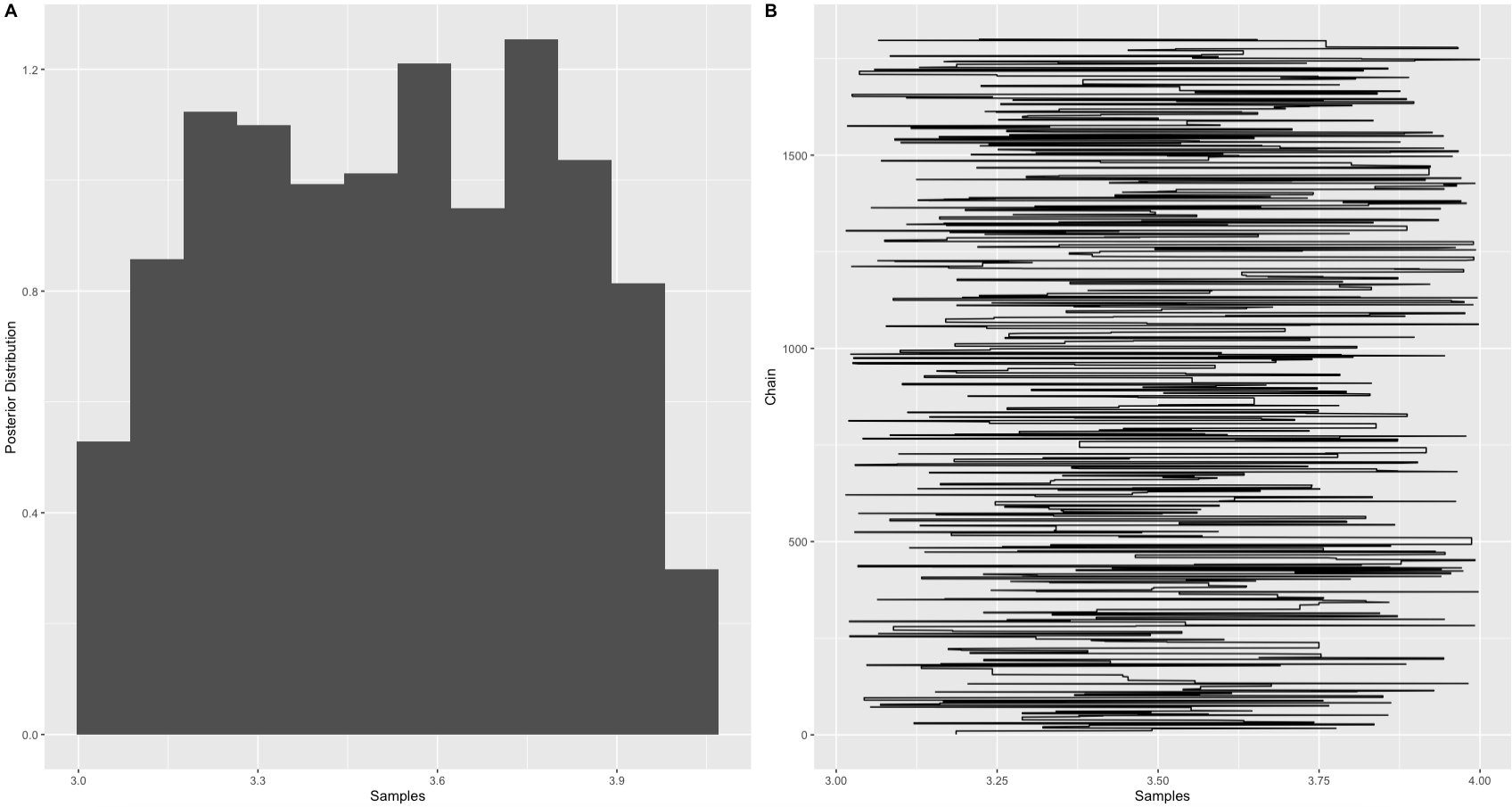}
\caption{The posterior distribution and trace plot of the $\kappa$ based on MCMC samples from the Metropolis-Hastings algorithm using only the perturbed  DE variable $b_1(z)$ to accept or reject the transitions under the Uniform proposal distribution.}   \label{fig_b1_u}
\end{figure}

\begin{figure}
\centering
\includegraphics[width=1\textwidth]{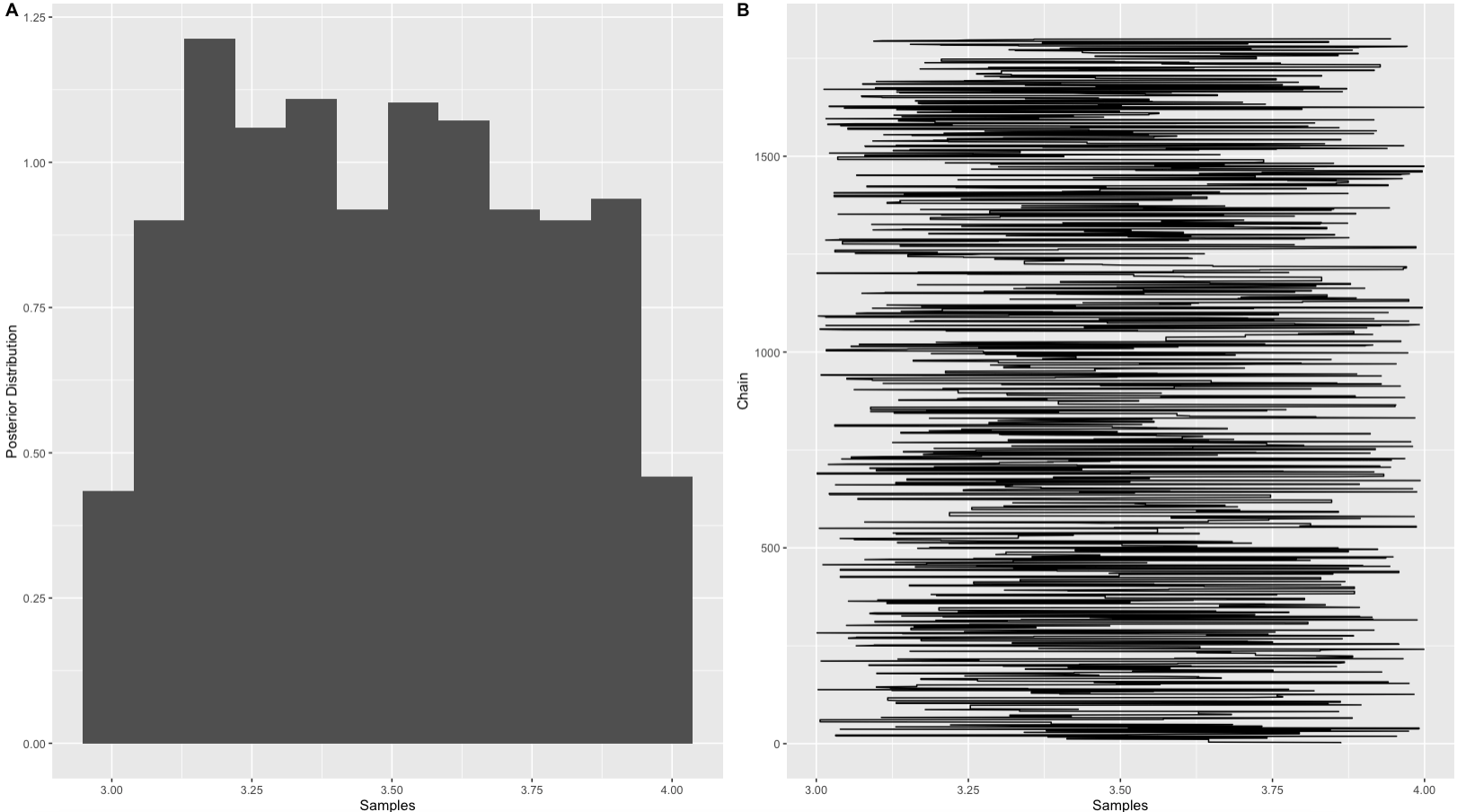}
\caption{The posterior distribution and trace plot of the $\kappa$ based on MCMC samples from the Metropolis-Hastings algorithm using only the perturbed DE variable $f_{1m}(z)=g_m(z)$ to accept or reject the transitions under the Uniform proposal distribution.}   \label{fig_gm_u}
\end{figure}

\begin{figure}
\centering
\includegraphics[width=1\textwidth]{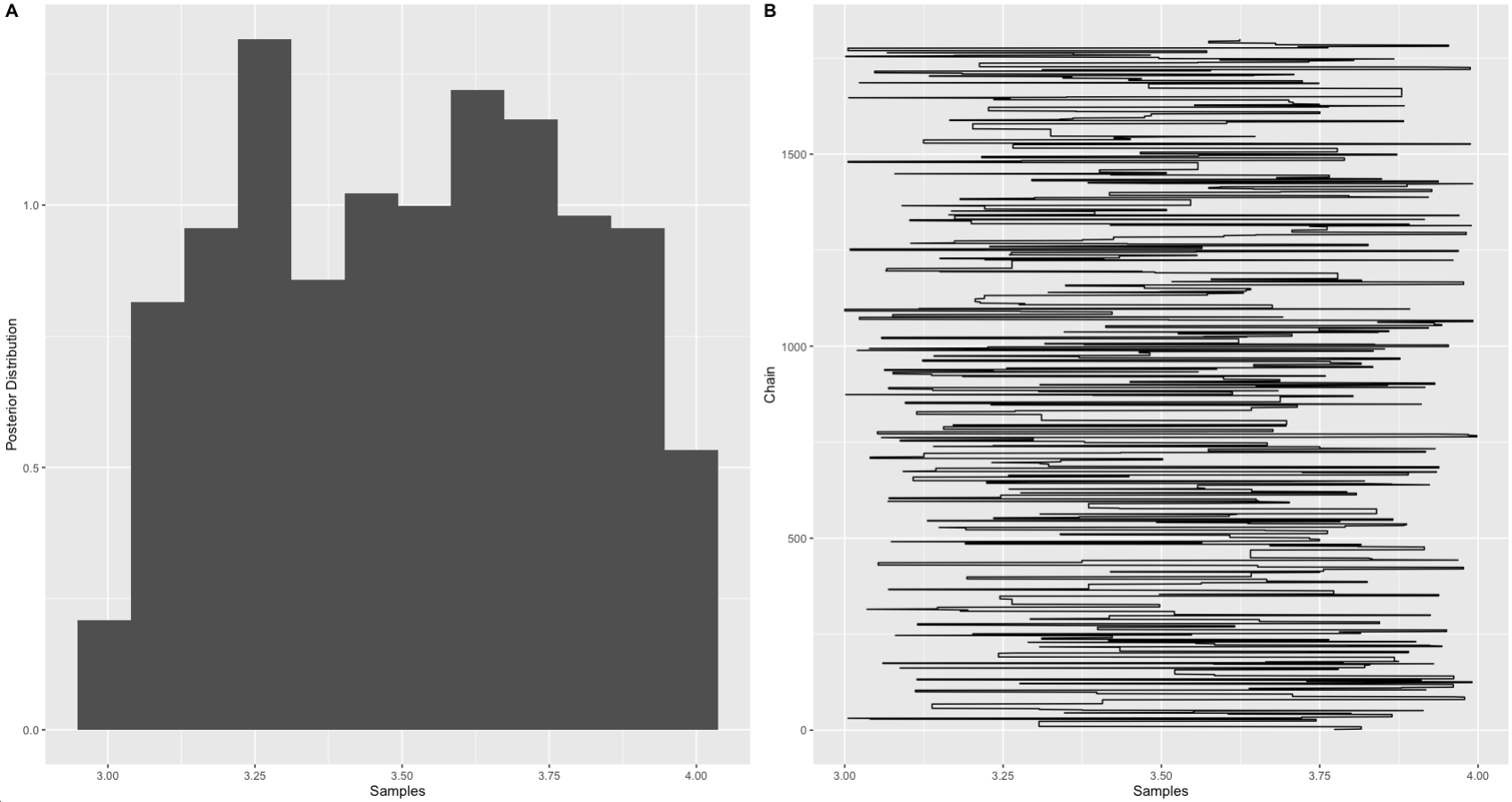}
\caption{The posterior distribution and trace plot of the $\kappa$ based on MCMC samples from the Metropolis-Hastings algorithm using both the perturbed DE variables $b_1(z)$ and $g_m(z)$  to accept or reject the transitions under the Uniform proposal distribution.}   \label{fig_b1_gm_u}
\end{figure}

\begin{figure}[ht]
\centering
\includegraphics[width=1\textwidth]{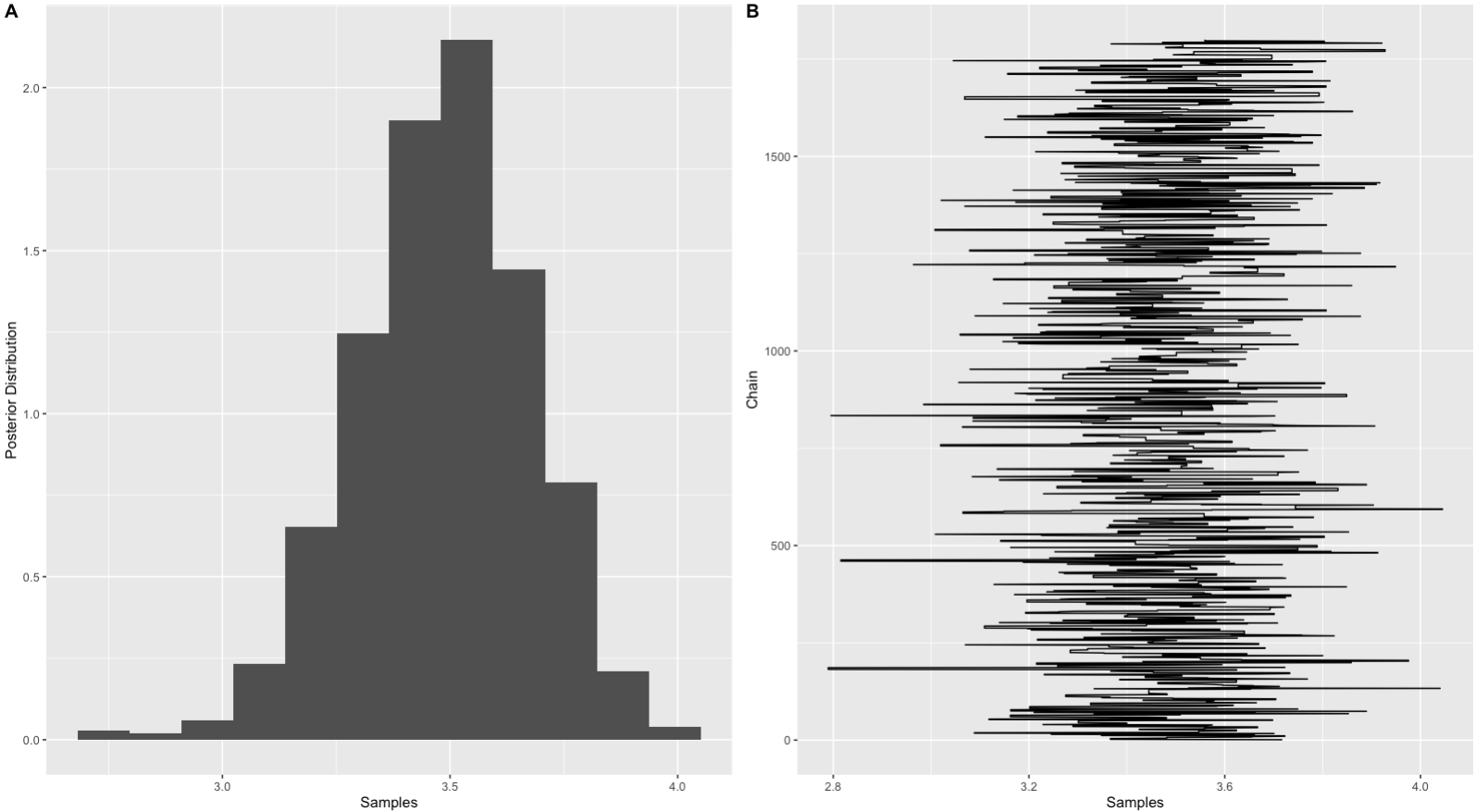}
\caption{The posterior distribution and trace plot of the $\kappa$ based on MCMC samples from the Metropolis-Hastings algorithm using only the perturbed  DE variable $b_1(z)$ to accept or reject the transitions under the Gaussian proposal distribution.}   \label{fig_b1_G}
\end{figure}

\begin{figure}[ht]
\centering
\includegraphics[width=1\textwidth]{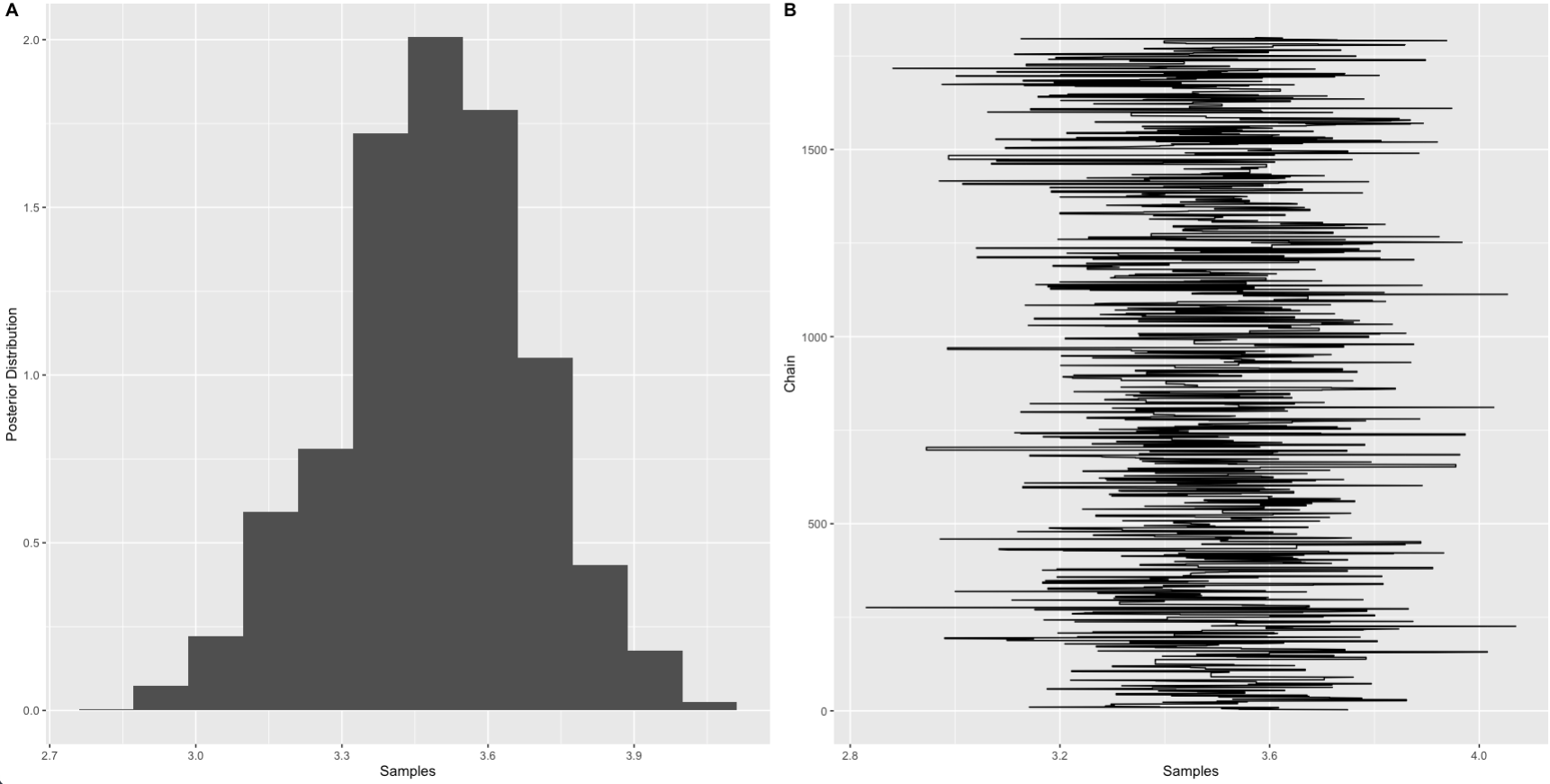}
\caption{The posterior distribution and trace plot of the $\kappa$ based on MCMC samples from the Metropolis-Hastings algorithm using only the perturbed DE variable $f_{1m}(z)=g_m(z)$ to accept or reject the transitions under the Gaussian proposal distribution.}   \label{fig_gm_G}
\end{figure}

\begin{figure}[ht]
\centering
\includegraphics[width=1\textwidth]{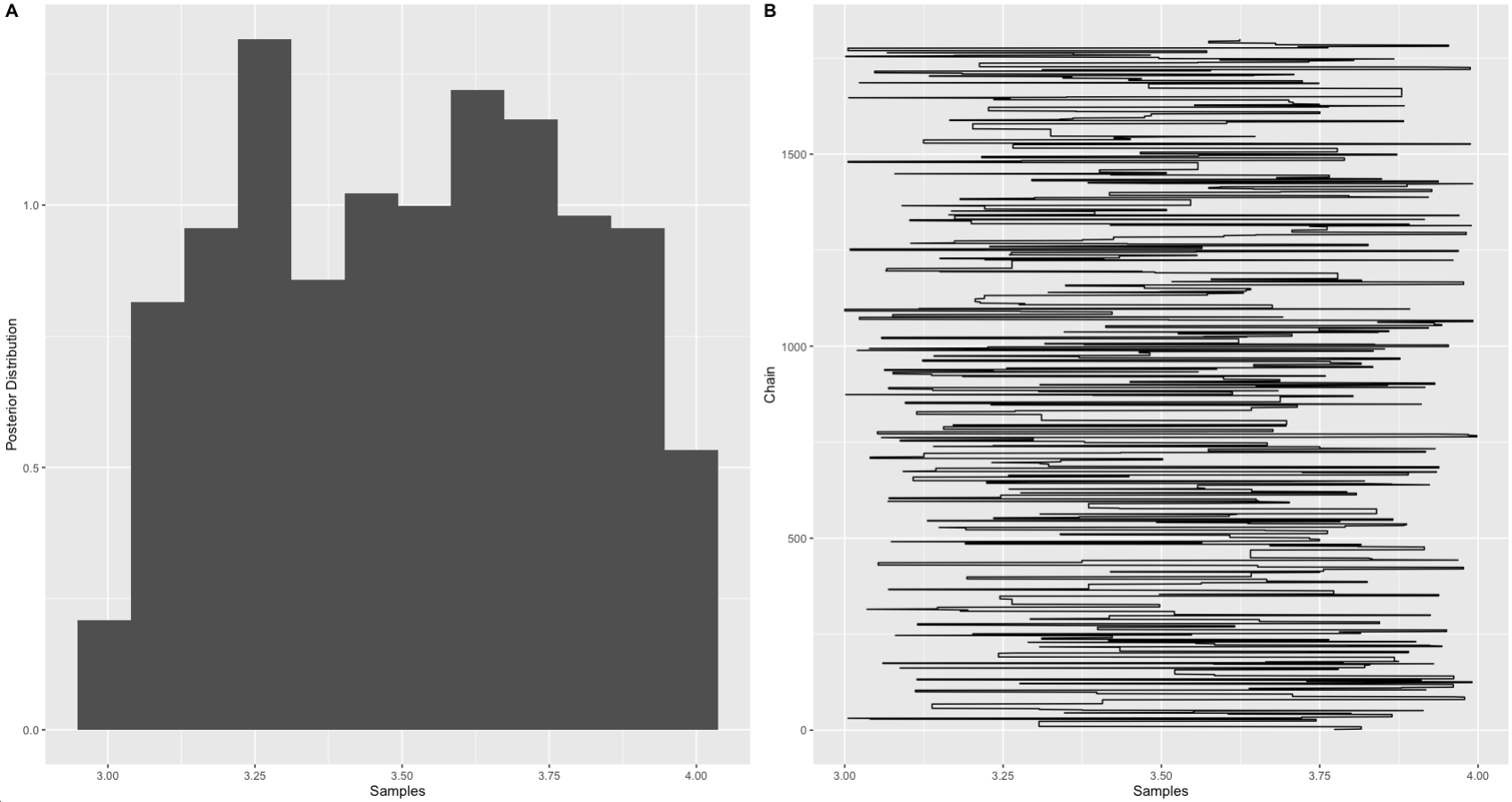}
\caption{The posterior distribution and trace plot of the $\kappa$ based on MCMC samples from the Metropolis-Hastings algorithm using both the perturbed DE variables $b_1(z)$ and $g_m(z)$ to accept or reject the transitions under the Gaussian proposal distribution.}   \label{fig_b1_gm_G}
\end{figure}

Figures \ref{fig_b1_u}-\ref{fig_b1_gm_G} show the posterior distributions and trace plots of the $\kappa$ parameter based on 1800 MCMC chains (after discarding the burn-in period) using our artificial neural networks-assisted Metropolis-Hastings approach. The results are based on three accept-reject approaches and two proposal distributions, including uniform and Gaussian distributions. 
From Figures \ref{fig_b1_u}-\ref{fig_b1_gm_G}, we observe that the distribution's posterior mean and posterior mode, as two Bayesian estimates of the $\kappa$ range between 3.2 and 3.8, which verifies the findings in the literature that $\kappa^* \approx 3.7858$ for the equations of motion in 4d elliptic class \cite{Antonelli:2019dqv}. From the trace plot, we easily observed that the MCMC chain easily searches the domain of the parameter of interest and supports the solution between 3.2 and 3.8. 

As one needs the ANNs in each iteration of the MCMC chain to evaluate the accept-reject steps, we show the results of the loss function of ANNs in Figures \ref{fig_l_b1_U}-\ref{fig_l_b1_gm_G} (in the Appendix). 
To show the convergence of the ANNs, we computed the difference between training and validation loss functions in the last epoch of the ANNs over 1800 MCMC samples. In each figure, we also show the entire trajectory of the training and validation loss functions over all the epochs for a randomly selected MCMC sample. 
From Figures \ref{fig_l_b1_U}-\ref{fig_l_b1_gm_G}, we clearly observe the loss differences are almost very small of magnitude $10^{-6}$ fluctuating around zero. This highlights the convergence of the ANNs in estimating the critical exponent in a Bayesian framework.

The unique solution for the elliptic case in four dimensions was achieved in \cite{Antonelli:2019dqv}. Indeed, the behaviour of $det A(k)$ near the last crossing of the
horizontal axis was estimated to be 
\begin{equation}
    \kappa^*_{4E} \approx 3.7858\,,
\end{equation}
which gives rise to a Choptuik exponent as  $\gamma_{4E} \approx 0.2641$. In this paper, we have used various statistical methods to actually explore the entire range of the critical exponent rather than exploring the last crossing. Surprisingly, our results 

\begin{equation}
  3.2  <\kappa^*_{4E} < 3.8
\end{equation}
is in perfect agreement with the literature, see \cite{Eardley:1995ns}. Hence, we can actually take this very non-trivial match as evidence that shows all our numerical sets up are reliable methods that can be extended to any dimension and  any matter content.


\section{Conclusion}\label{sec:con}

In this paper, we advance the current literature on critical exponents for black hole solutions by examining all solutions within the domains of the linear perturbation equations of motion. Specifically, we investigate the quantum perturbation theory for a four-dimensional Einstein-axion-dilaton system under an elliptic class of $\text{SL}(2,\mathbb{R})$ transformations. Utilizing quantum perturbation theory, we propose artificial neural network-assisted Metropolis-Hastings algorithms for Bayesian estimation of the critical exponent of the elliptic black hole solution in 4d.

Unlike conventional methods, we investigated the range of possible values for critical exponents using quantum perturbation theory. We conducted a thorough analysis not only of the perturbed differential equation (DE) variables $b_1(z)$ and $f_{1m}(z)=g_m(z)$ but also of the DE variables for both $b_1(z)$ and $g_m(z)$ equations simultaneously. This comprehensive approach allowed us to assess their effects on stochastic accept-reject transitions within the posterior distributions of the critical exponent. Our innovative probabilistic method stands out from existing techniques by offering the established solution while simultaneously exploring the entire spectrum of physically distinguishable critical exponents, accounting for potential numerical measurement errors. This advancement provides a more robust and inclusive understanding of critical exponents, highlighting the limitations of previous approaches and paving the way for more accurate predictions in complex systems.

Our new methods introduce innovative approaches for exploring the potential range of allowed values for the critical exponent, applicable to all different conjugacy classes of the $\sltr$
 transformation. Moreover, it is important to emphasize that these methods can be extended to various types of matter content. This is a direction we are keen to pursue in future research.

It is interesting to note that, besides the critical exponents' dependence on matter content, dimension, and various solutions of self-critical collapse ~\cite{ours}, these exponents can span an entire range of values rather than being confined to a single localized value. Therefore, we conclude that the conjecture regarding the universality of the Choptuik exponent does not hold. However, some universal behaviours might be embedded in the combinations of critical exponents and other parameters of the given theory that our current analysis has not accounted for. Our new findings clearly indicate that the standard expectations of statistical mechanics do not transfer to the context of critical gravitational collapse.


\section{Acknowledgments}

E. Hatefi would like to thank Pablo Diaz, E. Hirschmann, Philip Siegmann, L. Alvarez-Gaume, and A. Sagnotti for their useful discussions and support. E. Hatefi is supported by the María Zambrano Grant of the Ministry of Universities of Spain. Armin Hatefi acknowledges the support from the Natural Sciences and Engineering Research Council of Canada (NSERC).


\newpage
\section*{Appendix}

\begin{figure}[ht]
\centering
\includegraphics[width=1\textwidth]{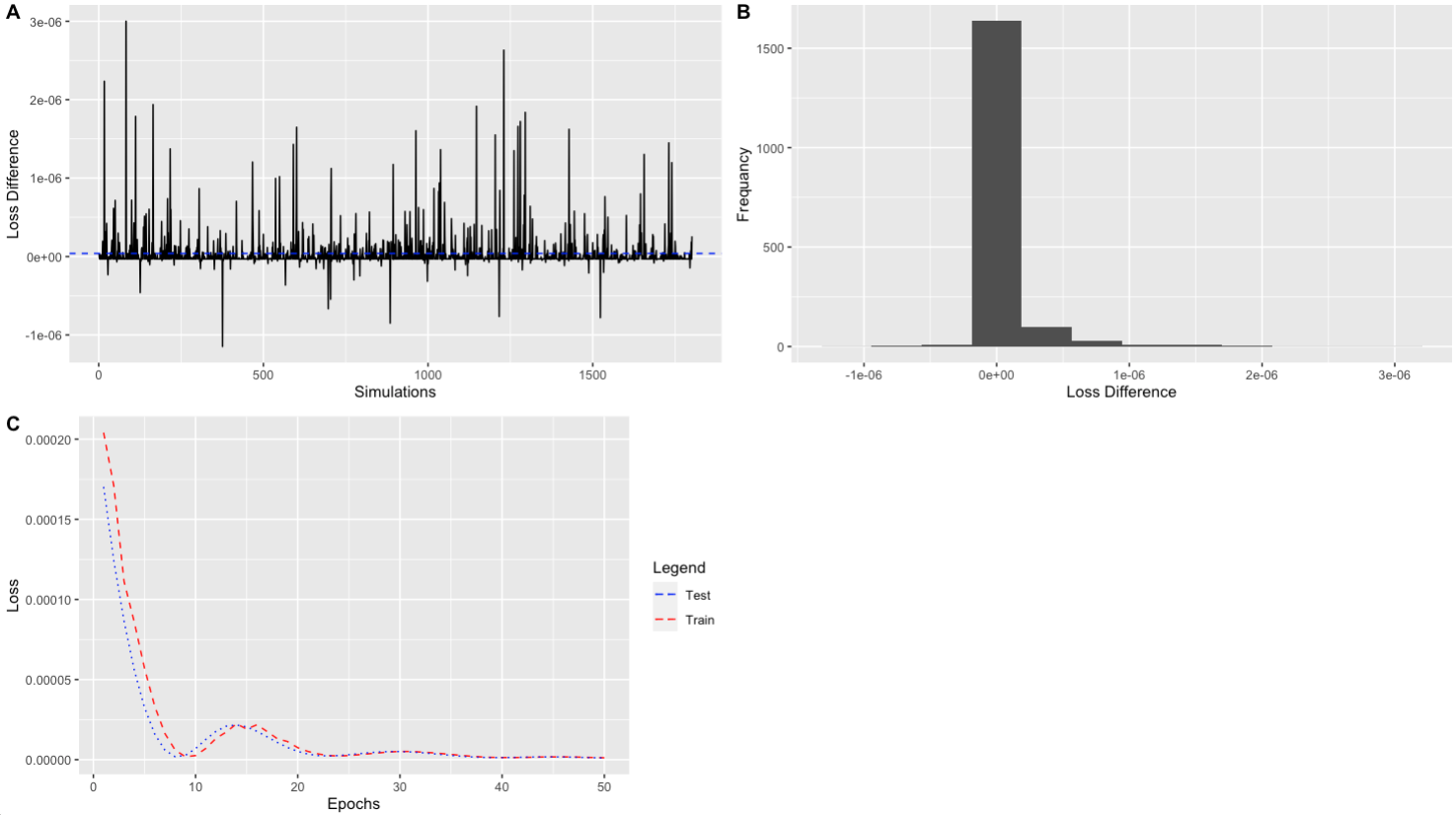}
\caption{ The difference between the train and validation losses in the last epochs over MCMC samples (A) and the histogram of the loss differences (B).  The train and validation losses for one random MCMC sampling (C). The MCMC samples are from the Metropolis-Hastings algorithm, using only the perturbed DE variable $b_1(z)$  to accept or reject the transitions under the Uniform proposal distribution.}   \label{fig_l_b1_U}
\end{figure}

\begin{figure}[ht]
\centering
\includegraphics[width=1\textwidth]{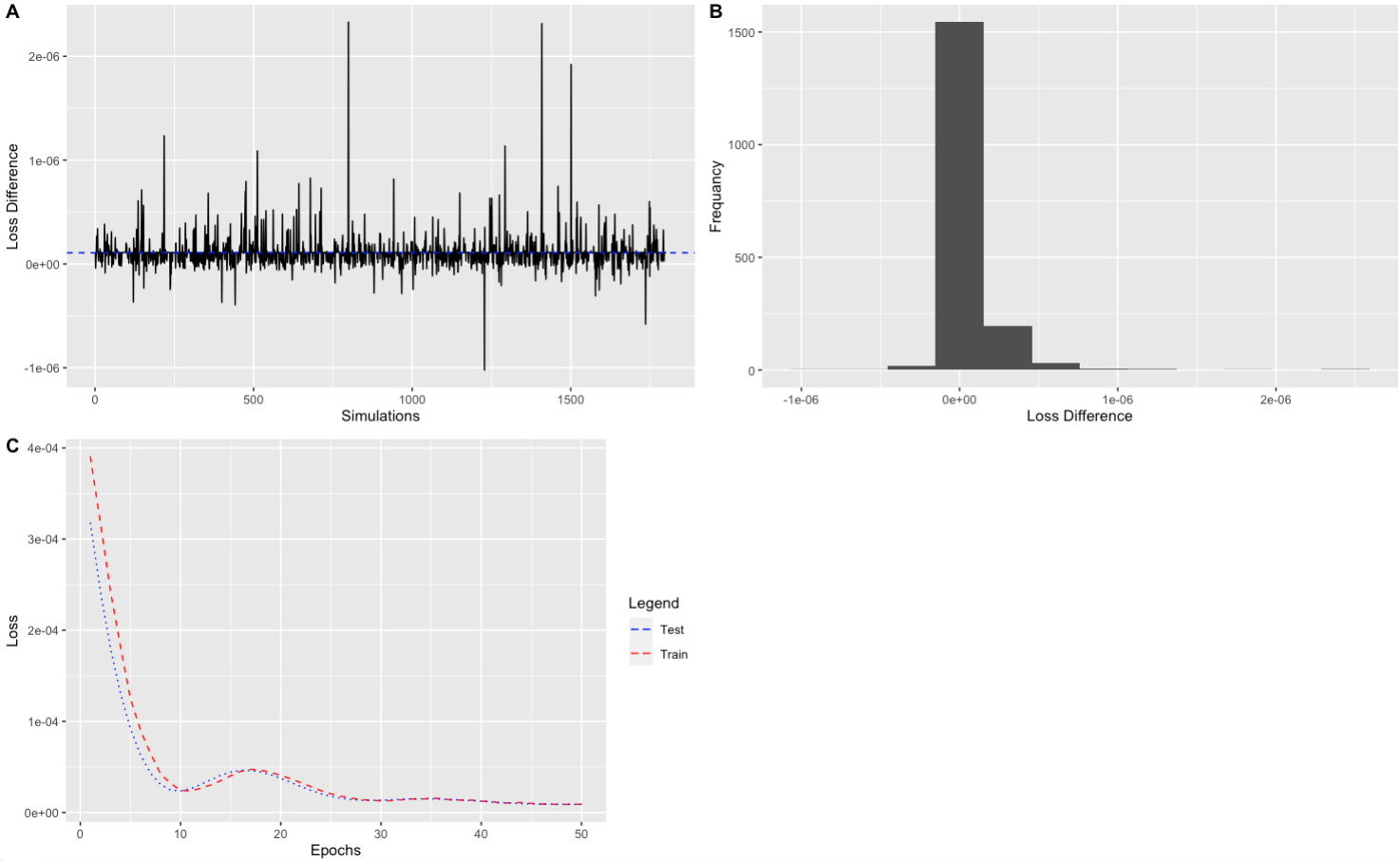}
\caption{ The difference between the train and validation losses in the last epochs over MCMC samples (A) and the histogram of the loss differences (B).  The train and validation losses for one random MCMC sampling (C). The MCMC samples are from the Metropolis-Hastings algorithm, using only the perturbed DE variable $g_m(z)$  to accept or reject the transitions under the Uniform proposal distribution.}   \label{fig_l_gm_U}
\end{figure}

\begin{figure}[ht]
\centering
\includegraphics[width=1\textwidth]{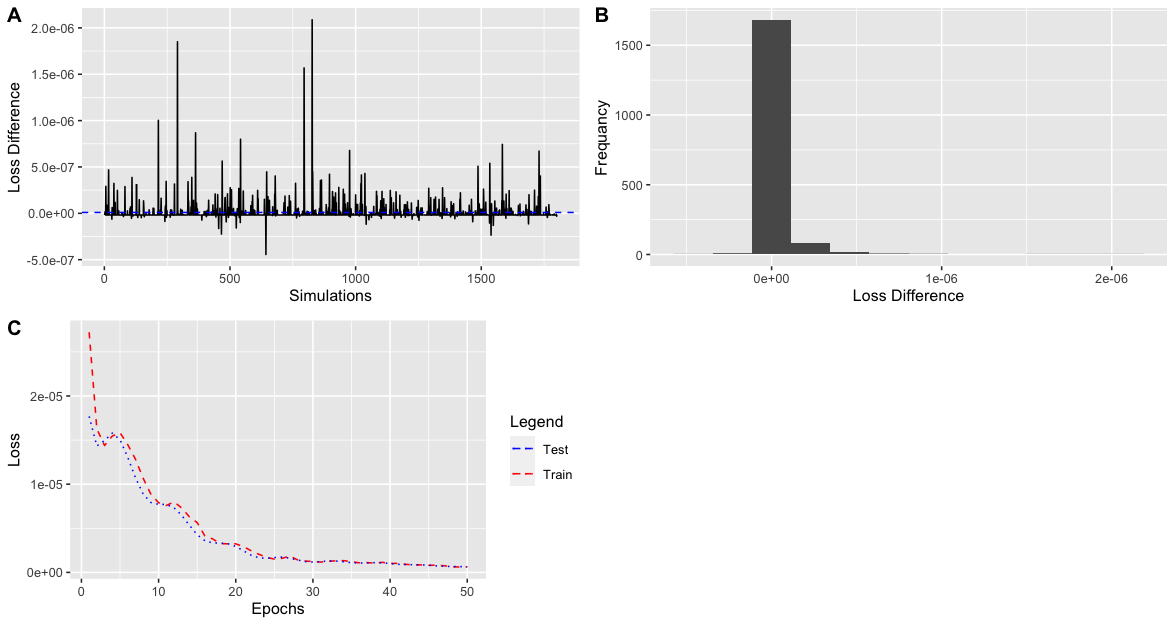}
\caption{ The difference between the train and validation losses in the last epochs over 1800 MCMC samples (A) and the histogram of the loss differences (B).  The train and validation losses for one random MCMC sampling (C). The MCMC samples are from the Metropolis-Hastings algorithm, using both perturbed DE variables $b_1(z)$ and $g_m(z)$  to accept or reject the transitions under the Uniform proposal distribution.}   \label{fig_l_b1_gm_U}
\end{figure}

\begin{figure}[ht]
\centering
\includegraphics[width=1\textwidth]{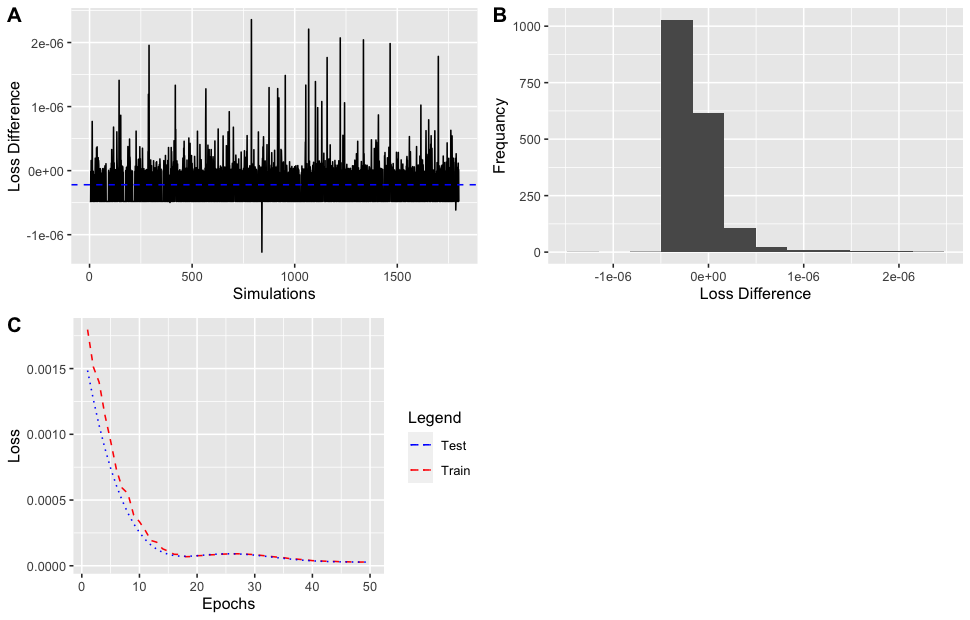}
\caption{ The difference between the train and validation losses in the last epochs over MCMC samples (A) and the histogram of the loss differences (B).  The train and validation losses for one random MCMC sampling (C). The MCMC samples are from the Metropolis-Hastings algorithm, using only perturbed DE variables $b_1(z)$  to accept or reject the transitions under the Gaussian proposal distribution.}   \label{fig_l_b1_G}
\end{figure}

\begin{figure}[ht]
\centering
\includegraphics[width=1\textwidth]{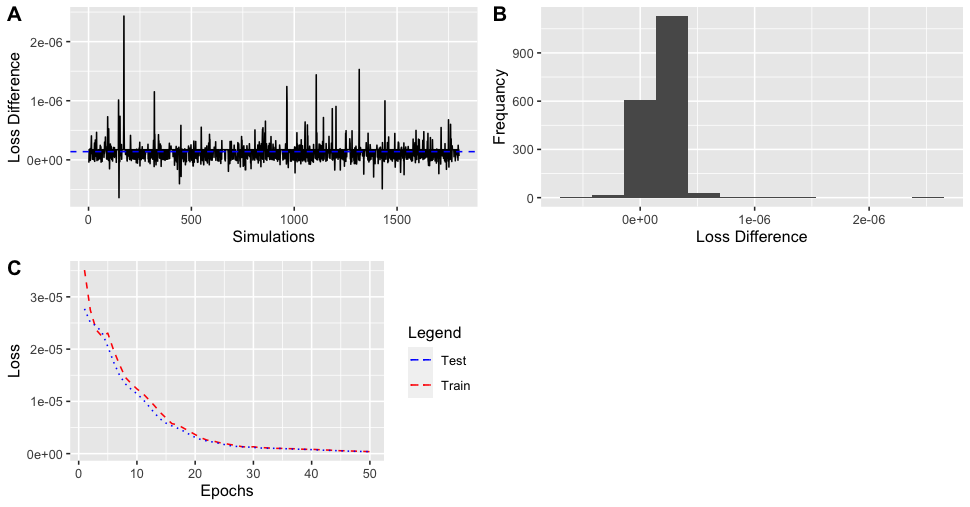}
\caption{The difference between the train and validation losses in the last epochs over MCMC samples (A) and the histogram of the loss differences (B).  The train and validation losses for one random MCMC sampling (C). The MCMC samples are from the Metropolis-Hastings algorithm, using only the perturbed DE variable $g_m(z)$  to accept or reject the transitions under the Gaussian proposal distribution.}   \label{fig_l_gm_G}
\end{figure}

\begin{figure}[ht]
\centering
\includegraphics[width=1\textwidth]{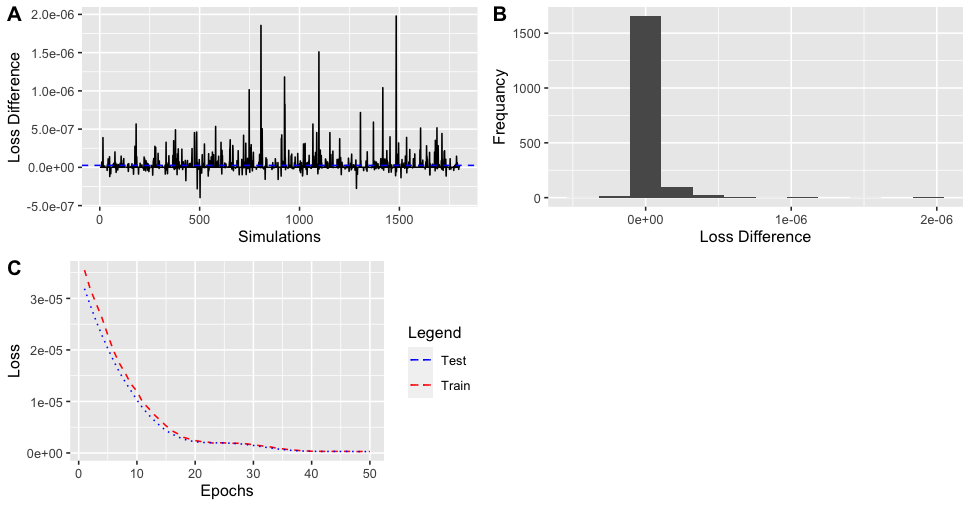}
\caption{ The difference between the train and validation losses in the last epochs over MCMC samples (A) and the histogram of the loss differences (B).  The train and validation losses for one random MCMC sampling (C). The MCMC samples are from the Metropolis-Hastings algorithm, using both perturbed DE variables $b_1(z)$ and $g_m(z)$  to accept or reject the transitions under the Gaussian proposal distribution.}   \label{fig_l_b1_gm_G}
\end{figure}


\end{document}